%% file: at_arxiv_main.tex
\documentclass[a4paper]{article}

\usepackage[english]{babel}
\usepackage[utf8]{inputenc}
\usepackage{lmodern}
\usepackage{microtype}

\usepackage{mathtools}
\usepackage{amsmath,amssymb,multirow}

\usepackage{amsthm}
\usepackage{bm}
\usepackage{subcaption}
\usepackage[svgnames]{xcolor}
\usepackage{siunitx}
\usepackage{booktabs}
\usepackage[normalem]{ulem}
\usepackage{graphicx}
\usepackage[width=1.2\columnwidth,textfont=small]{caption}

\usepackage{derivative} 
\usepackage{cases}      

\usepackage{enumitem}

\usepackage{pgfpages}
\pgfdeclarelayer{background}
\pgfdeclarelayer{foreground}
\pgfsetlayers{background,main,foreground}

\newtheorem{thm}{Theorem}

\newtheorem{rem}[thm]{Remark}
\newtheorem{assumption}[thm]{Assumption}

\newtheorem{cor}[thm]{Corollary}
\newtheorem{definition}[thm]{Definition}

\usepackage[colorlinks=true, allcolors=blue]{hyperref}

\title{\bf High-gain model-following control\\[0.5ex] for trajectory tracking}

\author{
	Niclas Tietze\thanks{Control Engineering Group,  Technische Universt\"at Ilmenau, P.O.~Box 10 05 65, D-98684 Ilmenau, Germany.} \and
	Kai Wulff$^*$\thanks{Corresponding author: \texttt{kai.wulff@tu-ilmenau.de}} \and 
	Johann Reger$^*$}

\begin{document}

\maketitle

\begin{abstract}
We consider trajectory tracking for minimum-phase nonlinear systems in Byrnes-Isidori form using the model-following control (MFC) architecture. 
The tracking problem is motivated by a hierarchical control concept where a higher-level instance provides the reference trajectory at run-time.
We present a computational efficient implementation of the feedback linearisation MFC design, and apply high-gain feedback in the process control loop (PCL) to achieve practical tracking in presence of Lipschitz perturbations. Our main results establish ultimate boundedness of the tracking error and give a constructive bound for the high-gain scaling parameter to achieve arbitrary tracking precision. Further we establish that the peaking phenomenon can be attenuated using MFC. We demonstrate the results via an automotive case study considering advanced engine-based cruise control.
\end{abstract}

\noindent\textbf{\textsf{Keywords:\hspace{0.5em}}}
Model-following control, nonlinear systems,
trajectory tracking, hierarchical control, peaking attenuation, engine-based cruise control


\section{Introduction}
The model-following control (MFC) architecture is a well-established two-degree-of-freedom control structure.
It consists of a process model that is simulated in the model control loop (MCL) and a process control loop (PCL) working on the actual process.
A typical configuration, which is a variation of the classical multi-loop structure by Horowitz \cite{Hor1963}, is depicted in Figure~\ref{fig:mfc_feedback_lin}.
The MCL can be interpreted as model-based feedforward \cite{WurK2009,Rop2009,Zei2012,AscSB2002} or reference generator \cite{NieM1998,SchDA2014}. 
That is, by simulating the closed MCL at run-time, we generate both a feedforward control signal and a state that is to be tracked in the PCL.
Typical designs of the MCL include inversion-based techniques \cite{DevCP1996,HagD2003a}, model predictive control \cite{SchAS2013,BauSS2014}, gain scheduling \cite{DwoB2013} or local model networks \cite{WilWR2018,WilWR2019}.

Motivated by the pioneering works of Tyler \cite{Tyl1964} and of Erzberger \cite{Erz1968}, MFC has been applied and studied in many variations and control applications \cite{WinR1970,TraB1978,Flo1981,Mcm1983,AlaKP1986}.
Due to the limitations in processing power MFC was initially applied to linear systems.
Design and analysis of the MFC are carried out in the frequency domain and state space, e.g. \cite{OsyK2010,DwoPD2009,SkoDP2003,SkoD2000,LiTH1998,Brz2011} and \cite{AmbCG1985,BroO1980,CheC1998,Rop1990}, respectively, and studies of MFC for multiple input multiple output systems can be found in \cite{DwoP2006,DwoPD2009,Sat2009} and \cite{Pre1972}.
The analyses show that the MFC architecture exhibits the capability to decouple the tracking performance and the robustness with respect to model uncertainties and disturbances.

In view of this, 
a large amount of research has been carried out on MFC for nonlinear systems.
On the one hand, MFC has been applied to nonlinear processes using a linear approximation of the nonlinear process in the MCL to carry over the methods from linear control theory.
For example, \cite{AmbCG1985} and \cite{SugO1993} apply Lyapunov redesign within the MFC architecture to stabilise a nominal linear process in presence of nonlinear perturbations.   
And \cite{Brz2012} applies linear MFC to the feedback linearised nonlinear process.
For both approaches, the nonlinearity of the MFC is due to the design of a nonlinear process controller.
On the other hand, the nonlinearity of the MFC can also be a result of a nonlinear model in the MCL.
Motivated by the case studies \cite{BauSS2014,HubGH2013,Pie2012,WilWR2019} our preliminary studies on high-gain MFC \cite{WilWR2022,WilWR2025} and \cite{TieWR2024CDCSMC,TieWR2024} apply MFC with a nonlinear model with feedback linearisation in both the MCL and the PCL.
The results suggest that the distinct robustness properties of the linear MFC can be carried over to the nonlinear case via feedback linearisation.
In particular, we show that high-gain feedback in PCL increases the robustness of the MFC compared to a conventional single-loop high-gain design \cite{WilWR2025}, while avoiding a well-known drawback of the high-gain feedback, the so-called peaking phenomenon, i.e. large (initial) control effort \cite{TieWR2024}.
See \cite{TsyP1999,YouKU1977}
and \cite{FraG1978,SusK1991}
for an introduction to high-gain design and in-depth investigations of the peaking phenomenon.

\textbf{Contribution:}
While our previous research on high-gain MFC consider set-point tracking, this study is dedicated to the
online trajectory tracking using high-gain control within the MFC scheme and its application within a case study.
In particular we consider the application of the MFC in the context of a hierarchical control scheme where the MFC serves as a low-level tracking controller, e.g. \cite{SchK2017}. 
In such scenario the reference trajectory provided by the higher-level instance is unknown prior to the process run-time and may exhibit inconsistencies to the initial process state.
Considering the perturbed nonlinear process in Byrnes-Isidori form, we introduce the feedback linearisation MFC design with a nonlinear model in the MCL.
We propose an efficient implementation of the nonlinear MFC requiring only the simulation of a linear model, which is equivalent to the nonlinear control law in the MCL, but requires less computational effort.
We show ultimate boundedness of the tracking error and give a constructive bound for the high-gain scaling parameter to achieve arbitrary tracking precision.
Applying high-gain feedback in the PCL, we establish the capability of the MFC to attenuate peaking by leveraging the initial state of the MCL, which is an additional degree of freedom of the architecture.
Finally, we demonstrate the results with an automotive case study, which is inspired by the results of 
\cite{ReiOW2020,ReiOW2020a} and \cite{LupDO2022,DegLO2023}.

\textbf{Structure:}
The next section defines the tracking problem considered.
In Section~\ref{sec:MFC} we introduce the proposed model-following control and its efficient implementation.
Section~\ref{sec:peaking} presents the high-gain design of the PCL and establishes the tracking and the peaking attenuation capabilities of the design.
Finally, in Section~\ref{sec:case_study} we present a case study of an engine-based cruise control.


\section{Problem Definition}\label{sec:propblem_definition}
We consider the nonlinear system in Byrnes-Isidori form
\begin{subequations} \label{eq:system}
\begin{align}
	\dot{\xi} &= A \, \xi + B \big(
	a(\xi,\eta) 
	+ b(\xi,\eta) \, u
	+ \Delta(\xi,\eta,t)
	\big),
	\label{eq:system_external}
	\\
	\dot{\eta} &= q(\xi,\eta),
	\label{eq:system_internal}
	\\
	y &= \xi_1,
	\label{eq:system_output}
\end{align}
\end{subequations}
where $\xi(t) = [\xi_1(t),...,\xi_{n_\xi}(t)]^\top \in \mathbb{R}^{n_\xi}$, $\xi(0) =\xi_0$ and $\eta(t) \in \mathbb{R}^{n_\eta}$, $\eta(0) = \eta_0$ denote the external and internal state, resp., and $u(t) \in \mathbb{R}$ is the input. 
The pair $(A,B)$ is in Brunovsk\'{y}-form and $n = n_\xi + n_\eta>0$ for $n_\eta \geq 0$.
Thus, relative degree w.r.t. the output $y$ is $n_\xi \geq 1$.
The internal dynamics \eqref{eq:system_internal} with locally Lipschitz right-hand side $q:\mathbb{R}^{n_\xi} \times \mathbb{R}^{n_\eta} \to \mathbb{R}^{n_\eta}$ are assumed to be input-to-state-stable (ISS) w.r.t. the input $\xi$.
\begin{definition}[\cite{Kha2002}]
The dynamics \eqref{eq:system_internal} are ISS w.r.t. the input $\xi$ if there exist a class $\mathcal{KL}$ function $\beta_\mathrm{s}$ and a class $\mathcal{K}$ function $\alpha_\mathrm{s}$, such that for each $\eta_0 \in \mathbb{R}^{n_\eta}$ and every bounded $\xi$, the solution $\eta$ satisfies
\begin{align*}
	\Vert \eta(t) \Vert_2 \leq \beta_\mathrm{s}(\Vert x_0 \Vert_2,t) \! + \! 	\alpha_\mathrm{s}\big(
	\sup_{0 \leq \tau \leq t} \Vert u(\tau) \Vert_2
	\big) \; \text{for all $t \geq 0$}.
\end{align*}
\end{definition}
Furthermore, the known functions $a:\mathbb{R}^{n_\xi} \times \mathbb{R}^{n_\eta} \to \mathbb{R}$ and $b:\mathbb{R}^{n_\xi} \times \mathbb{R}^{n_\eta} \to \mathbb{R}$ are continuous, and the function $\Delta: \mathbb{R}^{n_\xi} \times \mathbb{R}^{n_\eta} \times \mathbb{R} \to \mathbb{R}$, which represents an unknown perturbation, is piecewise continuous with respect to time and locally Lipschitz in $\xi$ and $\eta$.
We assume $|b(\xi,\eta)|\geq b_\mathrm{m}$ for all $(\xi,\eta) \in \mathbb{R}^{n_\xi} \times \mathbb{R}^{n_\eta}$ for some $b_\mathrm{m} > 0$.

The goal is to design feedback of $\xi$ and $\eta$ such that the output $y$ 
practically tracks the $n_\xi$ times continuously differentiable reference $y_\mathrm{d}$, which, together with its bounded time derivatives $y_\mathrm{d}, \, \dot{y}_\mathrm{d}, \, ..., \, y_\mathrm{d}^{(n_\xi)}$, is available at run-time only, but unknown prior to run-time.
We shall refer to this problem as \emph{online trajectory tracking}.
In particular, the control design shall enforce that the error between $\xi$ and the desired external state%
\begin{align}\label{eq:desired_state}
\xi_{\mathrm{d}} \coloneqq \begin{bmatrix}
y_\mathrm{d} & \dot{y}_\mathrm{d} & ... & y_\mathrm{d}^{(n_\xi-1)}
\end{bmatrix}^\top
\end{align}
is ultimately bounded as in Definition 4.6 of \cite{Kha2002}, i.e. there exists some $T_\infty > 0$ such that $	\Vert \xi(t) - \xi_{\mathrm{d}}(t) \Vert_2 \leq r_\infty$ for all $t \geq T_\infty$
with an arbitrarily small ultimate bound $r_\infty > 0$. 
Finally, we assume without loss of generality that $r_\mathrm{d} \geq 0$ is a strict bound of $\Vert \xi_{\mathrm{d}}\Vert_2$ such that \mbox{$\sup_{t \geq 0} \Vert \xi_{\mathrm{d}}(t) \Vert_2 < r_\mathrm{d}$}.

\begin{rem}\label{rem:hierarchical_control}
Note that we do not require knowledge of the reference $\xi_{\mathrm{d}}$ prior to run-time.
Thus, the definition of the tracking problem facilitates real-time trajectory generation, for example by a superimposed control loop as part of a hierarchical control architecture.
In particular this involves that the reference may exhibit an inconsistent initial value $\xi_{\mathrm{d}}(0)\neq \xi_0$, which is subject of our peaking analysis in Section~\ref{sec:peaking_attenuation}.
\end{rem}

\begin{rem}
The special case of a constant reference $y_\mathrm{d}$ corresponds to set-point tracking, with $y_\mathrm{d} \equiv 0$ yielding stabilisation of the origin, \cite{WilWR2025,TieWR2025}.
In case $n_\xi = n$, the internal dynamics \eqref{eq:system_internal} are dropped and the functions $a$, $b$ and $\Delta$ only depend on $\xi$ and $(\xi,t)$, respectively.
Furthermore, even tough we introduce \eqref{eq:system} and \eqref{eq:perturbation_assumption} globally, similar results can be obtained locally. 
\end{rem}

\section{Model-Following Control}\label{sec:MFC}
\begin{figure*}[t]
\centering
\includegraphics[width=\columnwidth]{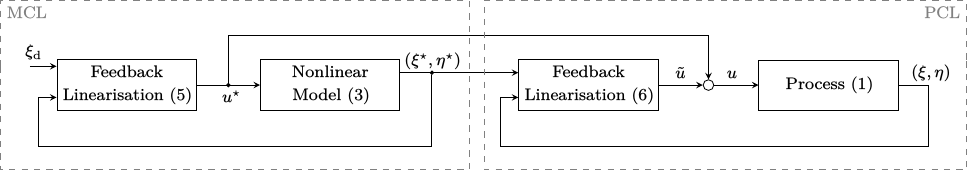}
\caption{Block diagram of the MFC architecture with nonlinear process model and feedback linearisation in both the MCL and the PCL.}
\label{fig:mfc_feedback_lin}
\end{figure*}

This section introduces the MFC architecture.
Generalising the results of \cite{WilWR2025} from set-point to trajectory tracking, we first present the conventional feedback linearisation design with a nonlinear model.
Then, we leverage the idea of \cite{TieWR2025ECC} to propose a mathematically equivalent control law, which requires only simulation of a linear model.
The two implementations are illustrated in Figure~\ref{fig:mfc_feedback_lin} and Figure~\ref{fig:mfc_feedback_lin_linear_model}.

\subsection{Design with Nonlinear Model}
The MCL uses the nominal model 
\begin{subequations}\label{eq:nonlinear_model}
\begin{align}
\dot{\xi}^\star &= A \, \xi^\star + B \big(
a(\xi^\star,\eta^\star) + b(\xi^\star,\eta^\star)\, u^\star 
\big),
\\
\dot{\eta}^\star &= q(\xi^\star,\eta^\star),
\end{align}
\end{subequations}
with states $\xi^\star(t) \in \mathbb{R}^{n_\xi}$, $\eta^\star(t) \in \mathbb{R}^{n_\eta}$ and input $u^\star(t) \in \mathbb{R}$. 
Define the error states $\tilde{\xi} = \xi - \xi^\star$ and $\tilde{\eta} = \eta - \eta^\star$, and decompose the control input $u = u^\star + \tilde{u}$ into the inputs $u^\star$ and $\tilde{u}$ of the MCL and the PCL, respectively.
The dynamics of the PCL can be written as
\begin{subequations}
\begin{align*}
\dot{\tilde{\xi}} &= A \, \tilde{\xi} + B \big(
\tilde{a}(\xi^\star,\eta^\star,\xi,\eta,u^\star)
+ b(\xi,\eta) \, \tilde{u}
+ \Delta(\xi,\eta,t)
\big),
\\
\dot{\tilde{\eta}} &= q(\xi,\eta) - q(\xi^\star,\eta^\star)
\end{align*}
\end{subequations}
with the auxiliary nonlinearity
\begin{align*}
\tilde{a}(\xi^\star\! \! ,\eta^\star\! \!,\xi,\eta,u^\star) \! = \! a(\xi,\eta) \!-\! a(\xi^\star \! ,\eta^\star) 
\!+\! \big(
b(\xi,\eta) \!-\! b(\xi^\star\!,\eta^\star)
\big) u^\star \! .
\end{align*}
Introducing the auxiliary inputs $v^\star(t) \in \mathbb{R}$ and $\tilde{v}(t) \in \mathbb{R}$, we apply the feedback linearisation control~laws 
\begin{gather}
u^\star = \frac{-a(\xi^\star,\eta^\star) + y_\mathrm{d}^{(n_\xi)} + v^\star}{b(\xi^\star,\eta^\star)}
\label{eq:u_mfc_feedback_linearisation_mcl}
\shortintertext{and}
\tilde{u} = \frac{-\tilde{a}(\xi^\star,\eta^\star,\xi,\eta,u^\star) + \tilde{v}}{b(\xi,\eta)}
\label{eq:u_mfc_feedback_linearisation_pcl}
\end{gather}
to the MCL and the PCL, respectively.
Define the model tracking error $\tilde{\xi}^\star = \xi^\star -
\xi_{\mathrm{d}}$ and note that $\dot{\xi}_\mathrm{d} = A \, \xi_{\mathrm{d}} + B \, y_\mathrm{d}^{(n_\xi)}$.
The dynamics of the open-loop system can then be written as
\begin{subequations}\label{eq:open_loop_nonlinear_model}
\begin{align}
\dot{\tilde{\xi}}^\star &= A \, \tilde{\xi}^\star + B \, v^\star,
\label{eq:open_loop_nonlinear_model_mcl_extern}
\\
\dot{\eta}^\star &= q(\xi^\star,\eta^\star),
\\
\dot{\tilde{\xi}}\phantom{^\star} &= A \, \tilde{\xi} + B \big(
\tilde{v} + \Delta(\xi,\eta,t)
\big),
\label{eq:open_loop_nonlinear_model_pcl_extern}
\\
\dot{\tilde{\eta}}\phantom{^\star} &= q(\xi,\eta) - q(\xi^\star,\eta^\star).
\label{eq:open_loop_nonlinear_model_pcl_intern}
\end{align}
\end{subequations}
We refrain from substitution $\xi = \xi^\star + \tilde{\xi} = \tilde{\xi}^\star + \xi_{\mathrm{d}} + \tilde{\xi}$ and $\eta = \eta^\star + \tilde{\eta}$ for brevity of the expressions. 
For the nominal case $\Delta(\xi,\eta,t) \equiv 0$, we choose $v^\star$ and $\tilde{\xi}$ such that the origins of \eqref{eq:open_loop_nonlinear_model_mcl_extern} and \eqref{eq:open_loop_nonlinear_model_pcl_extern} are asymptotically stable so that the tracking error
\begin{align}\label{eq:tracking_error}
\xi - \xi_{\mathrm{d}} 
= \xi^\star + \tilde{\xi}  - \xi_{\mathrm{d}}
= \tilde{\xi}^\star + \tilde{\xi}
\end{align}
asymptotically converges to zero, i.e. the MFC guarantees asymptotic tracking of $\xi_{\mathrm{d}}$
Then, the internal states $\eta$ and $\eta^\star$ of the input-to-state-stable internal dynamics of the process and the model are bounded.

\newpage
\subsection{Efficient Implementation}
\begin{figure*}[t]
\centering
\includegraphics[width=\columnwidth]{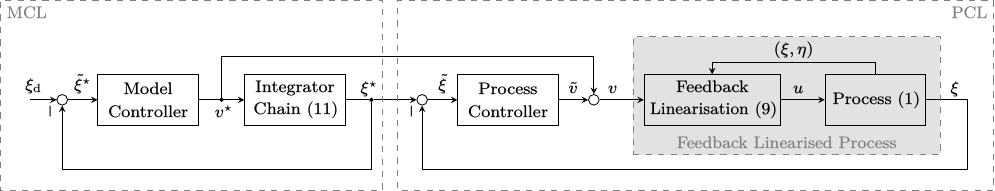}
\caption{Block diagram of the equivalent nonlinear MFC architecture with more efficient implementation.}
\label{fig:mfc_feedback_lin_linear_model}
\end{figure*}
Applying the feedback linearisation control~law
\begin{align}\label{eq:u_feedback_linearisation}
u = \frac{-a(\xi,\eta) + y_\mathrm{d}^{(n_\xi)} +v}{b(\xi,\eta)}
\quad \text{ with } \quad
v = v^\star + \tilde{v}
\end{align}
to the process \eqref{eq:system}, the dynamics of the open loop read
\begin{subequations}\label{eq:system_feedback_linearisation}
\begin{align}
\dot{\xi} &= A \, \xi + B \big(
y_\mathrm{d}^{(n_\xi)} + v
+ \Delta(\xi,\eta,t)
\big), 
\label{eq:system_feedback_linearisation_extern}
\\
\dot{\eta} &= q(\xi,\eta).
\end{align}
\end{subequations}
A nominal model of \eqref{eq:system_feedback_linearisation_extern} is the integrator chain
\begin{align}\label{eq:linear_model}
\dot{\xi}^\star =  A \, \xi^\star + B \big(
y_\mathrm{d}^{(n_\xi)} + v^\star
\big),
\quad \xi^\star(0) = \xi^\star_0.
\end{align}
Moreover, the dynamics of the error $\tilde{\xi} = \xi - \xi^\star$ satisfy
\begin{align*}
\dot{\tilde{\xi}} = A \, \tilde{\xi} + B \big(
\tilde{v} + \Delta(\xi,\eta,t)
\big).
\end{align*}
The dynamics of the open-loop MFC system,
\begin{subequations}\label{eq:open_loop}
\begin{align}
\dot{\tilde{\xi}}^\star &= A \, \tilde{\xi}^\star + B \, v^\star
\label{eq:open_loop_mcl}
\\
\dot{\tilde{\xi}}\phantom{^\star} &= A \, \tilde{\xi} + B \big(
\tilde{v} + \Delta(\xi,\eta,t)
\big), 
\label{eq:open_loop_pcl_extern}
\\
\dot{\eta}\phantom{^\star} &= q(\xi,\eta)
\label{eq:open_loop_pcl_intern}
\end{align}
\end{subequations}
comprise the dynamics of the model tracking error $\tilde{\xi}^\star$ in the MCL, the dynamics of $\tilde{\xi}$ and the internal dynamics of the process in the PCL.
Again, for $\Delta(\xi,\eta,t) \equiv 0$, tracking is achieved by appropriate choice of $v^\star$ and~$\tilde{v}$.
\begin{rem}
Note that $y_\mathrm{d}^{(n_\xi)}$ is available at run-time by assumption.
Thus the dynamics of the MCL \eqref{eq:linear_model} can be simulated at run-time, although they include the reference~$y_\mathrm{d}^{(n_\xi)}$.
\end{rem}

\subsection{Comparison of the Implementations}
The MFC \eqref{eq:open_loop_nonlinear_model} with the nonlinear model is illustrated in Figure~\ref{fig:mfc_feedback_lin}.
In the MCL, we obtain the linear external dynamics \eqref{eq:open_loop_nonlinear_model_mcl_extern} of the model tracking error by applying the feedback linearisation control law \eqref{eq:u_mfc_feedback_linearisation_mcl} to the nonlinear model \eqref{eq:nonlinear_model} of the process.
And applying $u = u^\star + \tilde{u}$ with feedback linearisation $\tilde{u}$ in~\eqref{eq:u_mfc_feedback_linearisation_pcl} to the process yields the external dynamics \eqref{eq:open_loop_nonlinear_model_pcl_extern} of the PCL.
Doing so, the design requires simulation of both the internal and the external dynamics of the nonlinear process of order $n = n_\xi + n_\eta$.

Figure~\ref{fig:mfc_feedback_lin_linear_model} depicts the MFC implementation \eqref{eq:open_loop}.
The MCL comprises the linear integrator chain \eqref{eq:linear_model} of order $n_\xi$, and feedback linearisation~\eqref{eq:u_feedback_linearisation} is applied to the PCL.

Comparing \eqref{eq:open_loop_nonlinear_model} and \eqref{eq:open_loop}, we see that both implementations yield identical error dynamics for $\tilde{\xi}^\star$ and $\tilde{\xi}$, respectively.
In particular, by substituting the auxiliary nonlinearity $\tilde{a}$ into \eqref{eq:u_mfc_feedback_linearisation_pcl}, it is readily verified that the control law $u = u^\star + \tilde{u}$ with the nonlinear model is equivalent to the control law \eqref{eq:u_feedback_linearisation} with the linear model.   
However, rather than simulating the copy \eqref{eq:nonlinear_model} of the nonlinear dynamics of the process of order $n = n_\xi + n_\eta$, the MFC in Figure~\ref{fig:mfc_feedback_lin_linear_model} requires only simulation of an $n_\xi$-th order integrator chain. 
Such an implementation offers advantages for both the simulation at run-time and the theoretical analysis.
At run-time, the simulation is computationally less demanding, in particular for systems with relative degree $n_\xi < n$, and it avoids numerical difficulties resulting from solving a differential equation containing the nonlinearities $a$ and~$b$.
For the theoretical analysis, we can exploit the ISS of the internal dynamics \eqref{eq:system_internal} of the process directly, rather than analysing the dynamics \eqref{eq:open_loop_nonlinear_model_pcl_intern} of $\tilde{\eta} = \eta - \eta^\star$. 

\begin{rem}
The difference between both implementations lies in the order of implementing the feedback linearisation control and the MFC scheme. 
By applying the MFC first, we consider a model of the nonlinear dynamics of the process, which we then linearise.
This approach leverages the original idea of the MFC for linear systems by Erzberger~\cite{Erz1968} in the sense that it simulates an exact copy of the process.
In contrast, by applying the feedback linearisation first, we obtain linear external dynamics that can be copied by a linear model in the MCL.
The design is facilitated by the interpretation of the MCL as a trajectory generator, which is considered in \cite{NieM1998,SchDA2014}.
In particular, the focus lies on the purpose of the MCL to generate a feedforward $v^\star$ and the solution $\xi^\star$, which is to be tracked.
In our case study in Section \ref{sec:case_study}, the implementation \eqref{eq:linear_model} reduces computational effort by approximately 40\%.
\end{rem}

\section{High-Gain Feedback in the PCL}\label{sec:peaking}
Considering the efficient implementation \eqref{eq:open_loop_mcl} of the MFC, we apply high-gain feedback in the PCL to achieve tracking with perturbations $\Delta$ that are globally Lipschitz in~$\xi$.

\begin{assumption}\label{ass:perturbation bound}
Let $L_\Delta, \delta \geq 0$ be known scalars such~that
\begin{align}\label{eq:perturbation_assumption}
\big|\Delta(\xi,\eta,t)\big| \leq \delta + L_\Delta \, \Vert \xi \Vert_2
\end{align}
for all $(\xi,\eta,t) \in \mathbb{R}^{n_\xi} \times  \mathbb{R}^{n_\eta} \times [0,\infty)$.
\end{assumption}
We first present the high-gain control design, then we study the invariance properties of the closed loop to establish the tracking properties of the design.
As to be expected, we require large gains to guarantee better tracking precision in presence of the perturbation \eqref{eq:perturbation_assumption}.
This typically causes large peaks in the control signal of the conventional single-loop high-gain feedback, the so-called peaking phenomenon.
It turns out that peaking can be avoided for the MFC.  
Essentially the advantage of the MFC is to attenuate the peaking, which occurs when it is required to track references $\xi_{\mathrm{d}}$ whose initial value $\xi_{\mathrm{d}}(0)$ does not match the initial state $\xi_0$ of the process with good precision.
Such inconsistency is a typical effect within a hierarchical control scheme or set-point control.
Note however, in such scenario the initial process state $\xi_0$ is available for the low-level tracking controller devised as MFC.
The main idea is to use the MCL to generate a consistent reference $\xi^\star$ which guides the process towards reference $\xi_{\mathrm{d}}$, whose initial value $\xi_{\mathrm{d}}(0)$ may not be given by the initial state $\xi_0$ of the process.
In doing so, we avoid peaking since we apply only high-gain feedback of the state of the PCL, which is close to the origin by~design.

\subsection{Control Design and Tracking Precision}
Choose the unscaled gain $k = [k_1,k_2,...,k_{n_\xi}]^\top 
\in \mathbb{R}^{n_\xi}$ such that $A - B \, k^\top$ is Hurwitz.
Given some positive scaling $\varepsilon < 1$, consider the scaled gain
\begin{align}\label{eq:scaled_gain}
\tilde{k}_\varepsilon 
= \begin{bmatrix}
\tfrac{k_1}{\varepsilon^{n_\xi}} & \tfrac{k_2}{\varepsilon^{n_\xi -1}} & \dots & \tfrac{k_{n_\xi}}{\varepsilon}
\end{bmatrix}^\top.
\end{align} 
Apply the control law \eqref{eq:u_feedback_linearisation} with 
\begin{align}\label{eq:u_high_gain}
v^\star = -k^\top \, \tilde{\xi}^\star,
&&
\tilde{v} = - \tilde{k}^\top_\varepsilon \, \tilde{\xi}.
\end{align}
Given the strict upper bound $r_\mathrm{d}$ of $\Vert \xi_{\mathrm{d}} \Vert_2$, our first theorem considers the tracking precision of the MFC design.
\begin{thm}\label{thm:tracking}
Given the closed-loop MFC system \eqref{eq:system}, \eqref{eq:u_feedback_linearisation}, \eqref{eq:linear_model} with perturbation satisfying Assumption~\ref{ass:perturbation bound} and the trajectory $\xi_{\mathrm{d}}$ from \eqref{eq:desired_state}.
Choose $0 < \varepsilon < 1$ such that $\frac{1}{\varepsilon} > 2 \, \Vert P B \Vert_2 \, L_\Delta$, for $P \! = \! P^\top \!> \! 0$ with $( A \!-\! B k^\top)^\top P + P  (A \!-\! B k^\top) = -I$.
Then, the solution $(\xi,\eta,\xi^\star)$ is bounded for every initial state $(\xi_0,\eta_0,\xi_0^\star) \in \mathbb{R}^{n_\xi} \times \mathbb{R}^{n_\eta} \times \mathbb{R}^{n_\xi}$.
Moreover, given the desired tracking precision $r_\infty > 0$, choosing $\varepsilon$ such that
\begin{align}\label{eq:scaling_precision}
\frac{1}{\varepsilon} > 2 \, \Vert P B \Vert_2  \, \sqrt{
\tfrac{\lambda_{\mathrm{max}}(P)}{\lambda_{\mathrm{min}}(P)}
} \, \frac{\delta + L_\Delta \, r_\mathrm{d}}{r_\infty} + 2 \, \Vert P B \Vert_2 \, L_\Delta
\end{align}
enforces that
the tracking error $\xi - \xi_{\mathrm{d}}$ is ultimately bounded with ultimate bound $r_\infty$.
\end{thm}
\begin{proof}
The proof is given in the appendix.
\end{proof}
The theorem shows that the high-gain feedback MFC achieves arbitrarily good tracking precision $r_\infty$ for sufficiency small scaling $\varepsilon$, which is, in particular, independent of the initial state $\xi_0^\star$ of the model \eqref{eq:linear_model}.
Recalling the decomposition $\xi - \xi_{\mathrm{d}} = \tilde{\xi}^\star + \tilde{\xi}$ of the tracking error into the model tracking error and the state of the PCL, the main idea of the proof is to consider the invariance properties of the MCL and the PCL individually.
In the MCL, we achieve asymptotic tracking $\xi^\star \rightarrow \xi_{\mathrm{d}}$ since there are no perturbations, and for the PCL, the high-gain feedback enforces an arbitrarily small ultimate bound for the error $\tilde{\xi}$ between the model and the process.

\begin{rem}
In the proof of Theorem~\ref{thm:tracking}, we consider the Lyapunov function \eqref{eq:lyapunov_function} given by the solution of the Lyapunov equation for $A - B \, k^\top$.
Following the approach of \cite{Raj1998} and \cite{Roeb2016}, using the solution of a Riccatti equation might allow to incorporate further structural properties of the perturbation to achieve better estimates for the scaling parameter $\varepsilon$ in \eqref{eq:scaling_precision}.
\end{rem}

\subsection{Single-Loop and Peaking Attenuation}\label{sec:peaking_attenuation}
A single-loop high-gain feedback for system \eqref{eq:system} is 
\begin{align}\label{eq:u_single_loop}
u_\mathrm{sl} = \frac{-a(\xi,\eta) + y_\mathrm{d}^{(n_\xi)} - \tilde{k}^\top_\varepsilon \, (\xi - \xi_{\mathrm{d}})}{b(\xi,\eta)}.
\end{align} 
Note that we select the same gain $\tilde{k}_\varepsilon$ for the single-loop controller \eqref{eq:u_single_loop} and the process controller of the MFC design \eqref{eq:u_high_gain}.
It is readily verified that the behaviour of the closed single-loop \eqref{eq:system}, \eqref{eq:u_single_loop} is equivalent to the MFC system \eqref{eq:system}, \eqref{eq:u_feedback_linearisation}, \eqref{eq:linear_model} for the initial state $\xi_0^\star = \xi_{\mathrm{d}}(0)$ of the model.
Thus, by application of Theorem~\ref{thm:tracking}, the single-loop design guarantees the same tracking precision as the MFC.
\begin{cor}\label{cor:single_loop}
Consider the closed single-loop \eqref{eq:system}, \eqref{eq:u_single_loop} with perturbation satisfying Assumption~\ref{ass:perturbation bound} and the trajectory $\xi_{\mathrm{d}}$ from \eqref{eq:desired_state}.
Given some $r_\infty > 0$, choose $ 0 < \varepsilon < 1$ satisfying \eqref{eq:scaling_precision}.
Then, the solution $(\xi,\eta)$ is bounded and the tracking error $\xi - \xi_{\mathrm{d}}$ is ultimately bounded with ultimate bound~$r_\infty$.
\end{cor} 
Considering the single-loop as a baseline for comparison, our next result
establishes the capability of the MFC scheme to attenuate peaking, i.e. large
initial control effort due to references with inconsistent initial values $\xi_{\mathrm{d}}(0) \neq \xi_0$.
In particular, we compare the control effort for the single-loop and the MFC with exact model initialisation $\xi_0^\star = \xi_0$.
To facilitate the comparison, without loss of generality we consider initial conditions of the process satisfying
\begin{align*}
\big(
\xi_0, \eta_0
\big) \in \mathcal{D} = \mathcal{D}_{\xi_0} \times \mathcal{D}_{\eta_0},
\end{align*}
where $\mathcal{D}_{\xi_0} \subset \mathbb{R}^{n_\xi}$ and $\mathcal{D}_{\eta_0} \subset \mathbb{R}^{n_\eta}$ are arbitrary compact neighbourhoods of the initial desired state $\xi_{\mathrm{d}}(0)$ and the origin, respectively.
In particular, we show that for small~$\varepsilon$, the maximal control
effort of the MFC is guaranteed to be less than the maximal control effort of the single-loop.
\begin{thm}\label{thm:u_comparison}
Given the trajectory $\xi_{\mathrm{d}}$ from \eqref{eq:desired_state}, consider the MFC system \eqref{eq:system}, \eqref{eq:u_feedback_linearisation}, \eqref{eq:linear_model} with exact model initialisation $\xi_0^\star = \xi_0$ and the single-loop \eqref{eq:system}, \eqref{eq:u_single_loop}.
There exists $\bar{\varepsilon} > 0$ such that, for all $\varepsilon < \bar{\varepsilon}$,
\begin{align*}
\max_{(\xi_0,\eta_0) \in \mathcal{D}}  \Big(
\sup_{t \geq 0} |u(t)|
\Big)
< 
\max_{(\xi_0,\eta_0) \in \mathcal{D}}   \Big(
\sup_{t \geq 0}|u_\mathrm{sl}(t)| 
\Big).
\end{align*}
\end{thm}
\begin{proof}
The proof is given in the appendix.
\end{proof}
As we decrease the scaling $\varepsilon$ to achieve good tracking precision, both, the MFC and the single-loop require increasing control effort.
However, Theorem~\ref{thm:u_comparison} shows that the increase is less for the MFC.
Thus, the MFC is better suited for high-precision tracking of references with inconsistent initial values, i.e. small $r_\infty$ and $\xi_\mathrm{d}(0) \neq \xi_0$. 

The idea of the proof is to leverage the capability of the MFC to decouple the tracking performance and the robustness via exact model initialisation.
In particular, the initial value of the single-loop control signal,
\begin{align}\label{eq:u_sl_initial_value}
u_\mathrm{sl}(0) = \frac{
-a(\xi_0,\eta_0) + y_\mathrm{d}^{(n_\xi)}(0) 
- \tilde{k}_\varepsilon^\top  (\xi_0 - \xi_{\mathrm{d}}(0)) 
}{b(\xi_0,\eta_0)},
\end{align}
gets arbitrarily large in view of the high-gain feedback $\tilde{k}_\varepsilon^\top  (\xi_0 - \xi_{\mathrm{d}}(0))$ for some $\xi_0$ in the neighbourhood $\mathcal{D}_{\xi_0}$ of $\xi_{\mathrm{d}}(0)$, i.e we cannot avoid peaking of the single-loop control signal for references with inconsistent initial values $\xi_{\mathrm{d}}(0) \neq \xi_0$.
For the MFC, however, choosing $\xi_0^\star = \xi_0$ guarantees that the initial value 
\begin{align}\label{eq:u_mfc_initial_value}
u(0) = \frac{
-a(\xi_0,\eta_0) + y_\mathrm{d}^{(n_\xi)}(0) 
- k^\top (\xi_0 - \xi_{\mathrm{d}}(0)) 
}{b(\xi_0,\eta_0)}
\end{align}
does not depend on $\varepsilon$ since $\tilde{\xi}_0 = \xi_0 - \xi_0^\star = 0$ by design.
Moreover, the upper bound of the state $\Vert \tilde{\xi}(t) \Vert_2$, which is fed back with the high gain $\tilde{k}_\varepsilon$, decreases as $\varepsilon$ decreases.
Conceptually, the MCL generates a new consistent reference $\xi^\star$, which is tailored to the initial state of the process, and guides the process towards the reference $\xi_{\mathrm{d}}$, whose initial value $\xi_{\mathrm{d}}(0) \neq \xi_0$. 
In doing so, the MFC architecture and the high-gain feedback design complement each other in the sense that only the small relative error $\tilde{\xi} = \xi - \xi^\star$ is fed back with the high gain, rather than the absolute tracking error~$\xi - \xi_{\mathrm{d}}$.%

\begin{rem}
Theorem~\ref{thm:u_comparison} generalises the result on peaking attenuation in \cite{TieWR2024} in several regards.
Firstly, the set-point tracking considered in \cite{TieWR2024} is extended to trajectory tracking.
Moreover, \cite{TieWR2024} does not consider any perturbation $\Delta$  in \eqref{eq:system}, whereas $b(\xi,\eta) \equiv 1$ and $a(\xi,\eta)$ is required globally Lipschitz.
However, \cite{TieWR2024} studies the so-called partial state-feedback, i.e. the case where only the external state $\xi$ is available for control and $\eta$ is unknown.
As a result the nonlinearity $a(\xi,\eta)$ cannot be fully compensated by the feedback and introduces an according perturbation.
\end{rem}

\begin{rem}[Set-point tracking]\label{rem:setpoint}
The set-point tracking is a special case of the problem considered, where 
$\xi_{\mathrm{d}}^\top \equiv \begin{bmatrix}
y_\mathrm{d}, 0,...,0
\end{bmatrix}$.
Except for trivial scenarios the set-point $\xi_{\mathrm{d}}$ is not equal to the initial process state $\xi_0$.
This setup emphasises the difference in the qualitative behaviour of the single-loop and the MFC for references with inconsistent initial values.
In the single-loop design the deviation from the set-point is fed back with high gain, and the speed of convergence to the set-point is determined by the desired tracking precision.
For the MFC, however, we implicitly solve the set-point tracking problem by tracking the non-constant reference $\xi^\star$, which steers $\xi$ to the set-point.  
See \cite{WilWR2022} and \cite{TieWR2025} for an in-depth study of the robustness properties and the decoupling of the precision and the performance of the high-gain MFC set-point tracking design, respectively.
\end{rem}

\section{Case Study: Cruise~Control}\label{sec:case_study}
To demonstrate the proposed MFC design and illustrate the results of the previous sections, we consider an automotive case study, namely an advanced engine-based cruise control.
In \cite{ReiOW2020,ReiOW2020a} and \cite{DegLO2023} feedback linearisation is applied to the so-called engine-based traction control.
Considering a car with a hybrid/electric engine, a model of the longitudinal vehicle dynamics including the power-train is considered.
The idea is to use the desired motor torque as the input for (indirect) traction control.
In particular, feedback linearisation with a PID controller is applied to control the desired motor torque such that the crankshaft speed tracks ramp-like reference trajectories during acceleration.  
In doing so, the wheel speed is controlled in an indirect manner to achieve optimal wheel slip.

The idea of the advanced engine-based cruise control in this contribution is to consider a state feedback rather than PID control for the auxiliary input of the feedback linearisation.
This facilitates advanced cruise control in the sense that arbitrary smooth references (for the wheel speed), which are demanded by a superimposed control system at run-time, can be tracked.
Note that the vehicle speed is close to the wheel speed for the typical cruise control scenario due to small slip.
The fact that the reference $\xi_{\mathrm{d}}$ is generated by the superimposed control system may cause a deviation of $\xi_{\mathrm{d}}(0)$ and $\xi_0$.   

\subsection{Powertrain Model}
\begin{figure}
\centering
\includegraphics{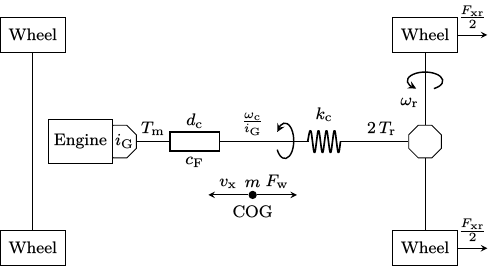}
\caption{Powertrain of a rear-wheel drive vehicle.}
\label{fig:example_model}
\end{figure}
We follow \cite{ReiOW2020a} to obtain the model for the longitudinal vehicle dynamics and the powertrain which is depicted in Figure~\ref{fig:example_model}.
The dynamics comprise the engine, the drivetrain, the rear axle including the tire dynamics, and the longitudinal vehicle acceleration.
For the simulation, we consider the specification of the rear-wheel drive vehicle from \cite{ReiOW2020a}.
The vehicle parameters are given in Table~\ref{table:vehicle_parameters}.
It shall be noted that we adapt the model by incorporating friction losses into the model of the drivetrain.
Moreover, even though we consider a rear-wheel drive vehicle, the design can also be applied to both front-wheel drive and all-wheel drive vehicles.

\textbf{Engine:}
For hybrid/electric vehicles, engine-based control design typically considers a linear approximation of the nonlinear engine dynamics~\cite{HorTT1998}.  
In the operating range of interest, the torque build-up of the hybrid/electric motor can be approximated by the first-order lag element%
\begin{align*}
\tau_\mathrm{m} \, \dot{T}_\mathrm{m} = T_{\mathrm{m},\mathrm{d}} - T_\mathrm{m}
\end{align*}
with the time-constant $\tau_\mathrm{m}$, the requested torque $T_{\mathrm{m},\mathrm{d}}$ and the torque $T_\mathrm{m}$ that is provided by the engine.

\textbf{Drivetrain:}
The dynamics of the crankshaft, the drive shaft and the half shafts are modelled as a torsion spring with twist angle $\phi_\mathrm{c}$, aggregated torsional stiffness $k_\mathrm{c}$ and damping factor $d_\mathrm{c}$.
In the operating range of interest, i.e. for rotational velocities of the crankshaft that correspond to cruise, a linear approximation of the losses due to rational friction (Coulomb and viscous) can be used.
Given the angular velocities $\omega_\mathrm{c}$ and $\omega_\mathrm{r}$ of the crankshaft and the rear axle, the dynamics read
\begin{subequations}
\begin{align*}
\dot{\phi}_\mathrm{c} &=  i_\mathrm{G}^{-1} \, \omega_\mathrm{c} - \omega_\mathrm{r},
\\
J_\mathrm{c} \, \dot{\omega}_\mathrm{c} &= T_\mathrm{m} - 2 \, i_\mathrm{G}^{-1} \, T_\mathrm{r} - c_\mathrm{F} \, \omega_\mathrm{c} ,
\\ 
T_\mathrm{r} &= k_\mathrm{c}\, \phi_\mathrm{c} + d_\mathrm{c} \left(
i_\mathrm{G}^{-1} \, \omega_\mathrm{c}-	\omega_\mathrm{r}
\right),
\end{align*}
\end{subequations}
where $i_\mathrm{G}$ is the total gear ratio from the motor to the rear axle and $c_\mathrm{F}$ is the coefficient of friction. 
The aggregated inertia of the drive side and the torque on the rear axle is  $J_\mathrm{c}$ and $2\, T_\mathrm{r}$, respectively.
In the context of cruise control we may consider the gear ratio as constant, i.e. a fixed gear, avoiding time-varying dynamics.

\textbf{Rear axle and wheels:}
We obtain the model of the rear axle by applying the law of angular momentum
\begin{align*}
J_\mathrm{r} \, \dot{\omega}_\mathrm{r} = 2\, T_\mathrm{r} - r_\mathrm{r} \, F_{\mathrm{xr}},
\end{align*}
where $J_\mathrm{r}$ is the inertia of the rear axle including wheels, $r_\mathrm{r}$ is the radius of the rear wheels and $F_{\mathrm{xr}}$ is the friction force between road and tire.
The friction force $F_{\mathrm{xr}}$ is modelled using the Pacejka's formula \cite{Pac2012} with the normal force $F_{\mathrm{zr}}$.
Given the tire parameters $B_\mathrm{r}$ and $C_\mathrm{r}$, we have
\begin{subequations}\label{eq:Magic_Formular}
\begin{align}
F_{\mathrm{xr}} &= \mu \, F_{\mathrm{zr}} \, \sin(C_\mathrm{r} \, \arctan(B_\mathrm{r} \, \lambda_{\mathrm{xr}})),
\\
F_{\mathrm{zr}} &= m \, g\, l_\mathrm{f} \left( l_\mathrm{f} + l_\mathrm{r} \right)^{-1},
\end{align}
\end{subequations}
where $g$ is the gravitational constant, $l_\mathrm{f}$ and $l_\mathrm{r}$  denote the distances from the centre of gravity (COG) of the vehicle to the front and the rear, respectively, and $\mu$ is the (time-variant) coefficient of friction between the tire and the road.
Moreover, given the vehicle velocity $v_\mathrm{x}$,
\begin{align}\label{eq:Slip_Coefficient_Rear}
\lambda_{\mathrm{xr}} = \frac{r_\mathrm{r}\, \omega_\mathrm{r} - v_\mathrm{x}}{v_{\mathrm{nr}}},
\quad
v_{\mathrm{nr}} = \mathrm{max}_\kappa \big\{
|r_\mathrm{r}\, \omega_\mathrm{r}|_\kappa, |v_\mathrm{x}|_\kappa
\big\},
\end{align} 
is a continuously differentiable approximation of the wheel slip $\lambda_{\mathrm{xr}}$ which is defined as the difference of the rotatory and the translatory motion of the wheel, normalised to the maximum of the absolute value of booth.
The operators $ \mathrm{max}_\kappa\left\{\cdot,\cdot\right\}$ and $|\cdot|_\kappa$ are defined as
\begin{subequations}
\begin{align*}
\mathrm{max}_\kappa\big\{ 
\gamma_1 , \gamma_2 
\big\} \! \coloneqq \tfrac{1}{2}\left(
\gamma_1 \! \! +  \! \gamma_2 \! + \! |\gamma_1 \! \!-\! \gamma_2|_\kappa 
\right),
\
|\gamma_1|_\kappa \! \coloneqq \! \sqrt{\gamma_1^2 \!+\! \kappa^2}
\end{align*}
\end{subequations} 
for $\gamma_1,\gamma_2 \in \mathbb{R}$ for some $\kappa > 0$, where the precision of the wheel slip approximation increases for $\kappa \rightarrow 0$.

\textbf{Longitudinal velocity:}
The longitudinal velocity $v_\mathrm{x}$ of the vehicle satisfies
\begin{align*} 
\dot{v}_\mathrm{x} = m^{-1}\big(
F_{\mathrm{xr}}-F_\mathrm{w}
\big),
\qquad 
F_\mathrm{w} = \tfrac{1}{2} \, \rho_\mathrm{a} \, c_\mathrm{w} \, A_{\mathrm{st}} \, v_\mathrm{x} \, |v_\mathrm{x}|, 
\end{align*}
where $F_\mathrm{w}$ is the aerodynamic drag force with air density $\rho_\mathrm{a}$, aerodynamic drag coefficient $c_\mathrm{w}$ and frontal area of the vehicle $A_{\mathrm{st}}$. 
Thus, the system dynamics read
\begin{subequations}\label{eq:example_model}
\begin{align}
\dot{T}_\mathrm{m} &=  \tau_\mathrm{m}^{-1} \big(
T_{\mathrm{m},\mathrm{d}} - T_\mathrm{m}
\big)
\\
\dot{\phi}_\mathrm{c} &=  i_\mathrm{G}^{-1} \, \omega_\mathrm{c} - \omega_\mathrm{r},
\label{eq:example_model_twist_angle}
\\
\dot{\omega}_\mathrm{c} &= 	J_\mathrm{c}^{-1} \big(
T_\mathrm{m} - 2 \, i_\mathrm{G}^{-1} \, T_\mathrm{r} - c_\mathrm{F} \, \omega_\mathrm{c}
\big),
\\
\dot{\omega}_\mathrm{r} &= J_\mathrm{r}^{-1}\big(
2 \, T_\mathrm{r} - r_\mathrm{r} \, F_{\mathrm{xr}}
\big),
\\ 
\dot{v}_\mathrm{x} &= m^{-1}(F_{\mathrm{xr}} - F_\mathrm{w}).
\end{align}
\end{subequations}
Choosing the requested motor torque as the input and the rotational velocity normalised to the wheel speed as the output, i.e. $u =T_{\mathrm{m,d}}$ and $y = i_\mathrm{G}^{-1} \, \omega_\mathrm{c}$, it is readily verified that $y$ has relative degree two with respect to the input~$u$.
Introducing the states
\begin{align*}
\xi = \begin{bmatrix}
\xi_ 1 \\ \xi_2
\end{bmatrix} = \begin{bmatrix}
y \\ \dot{y}
\end{bmatrix},
\qquad
\eta = \begin{bmatrix}
\eta_1 \\ \eta_2 \\ \eta_3
\end{bmatrix} = \begin{bmatrix}
\phi_\mathrm{c} \\ \omega_\mathrm{r} \\ v_\mathrm{x}
\end{bmatrix}\!,
\end{align*}
we obtain the dynamics in Byrnes-Isidori form
\begin{subequations}\label{eq:2WD_Byrnes}
\begin{align}
&\begin{aligned}
\dot{\xi} =  A \, \xi +  B  \big(
p_1 \,& \xi_1 +  p_2 \, \xi_2  + p_3 \, \eta_2 + b \, u 
\\
& 
-\tfrac{c_\mathrm{F}}{J_\mathrm{c}} (\tfrac{\xi_1}{\tau_\mathrm{m}} + \xi_2)
+ p_\mathrm{r}\, F_{\mathrm{xr}} 
+ p_\mathrm{c} \, \eta_1 
\big),
\end{aligned}
\label{eq:2WD_Byrnes_extern}
\\
&\dot{\eta} = q(\xi,\eta),
\label{eq:2WD_Byrnes_intern}
\end{align}
\end{subequations} 
where $A=\begin{bsmallmatrix}
0 & 1 \\ 0 & 0
\end{bsmallmatrix}$, $B = \begin{bsmallmatrix}
1 \\ 0
\end{bsmallmatrix}$, $b= (i_\mathrm{G} \, J_\mathrm{c} \, \tau_\mathrm{m})^{-1}$, 
\begin{gather*}
q(\xi,\eta) = \begin{bmatrix}
\xi_1 - \eta_2 \\
(2\, d_\mathrm{c}(\xi_1-\eta_2)+2\, k_\mathrm{c} \, \eta_1 - r_\mathrm{r} \, F_{\mathrm{xr}}) \, J_\mathrm{r}^{-1} \\
(F_{\mathrm{xr}}-F_\mathrm{w}) \, m^{-1}
\end{bmatrix},
\end{gather*}
and 
\vspace{-2ex}
\begin{align*}
p_1 &= \frac{(4\, d_\mathrm{c}^2 - 2\, J_\mathrm{r} \, k_\mathrm{c})\, \tau_\mathrm{m}-2\, J_\mathrm{r} \, d_\mathrm{c}}{i_\mathrm{G}^2 \, J_\mathrm{c}  \, J_\mathrm{r} \, \tau_\mathrm{m} },\\
p_2 &= -\frac{2\, J_\mathrm{r} \, \tau_\mathrm{m} \, d_\mathrm{c} + i_\mathrm{G}^2 \, J_\mathrm{c} \, J_\mathrm{r}}{i_\mathrm{G}^2 \, J_\mathrm{c}  \, J_\mathrm{r} \, \tau_\mathrm{m} },\\
p_3 &= \frac{(2\, J_\mathrm{r} \, k_\mathrm{c} - 4\, d_\mathrm{c}^2)\, \tau_\mathrm{m} + 2\, J_\mathrm{r} \, d_\mathrm{c}}{i_\mathrm{G}^2 \, J_\mathrm{c}  \, J_\mathrm{r} \, \tau_\mathrm{m} },
\\
p_\mathrm{c} &= \frac{4\, \tau_\mathrm{m} \, d_\mathrm{c} \, k_\mathrm{c} - 2\, J_\mathrm{r} \, k_\mathrm{c}}{i_\mathrm{G}^2 \, J_\mathrm{c}  \, J_\mathrm{r} \, \tau_\mathrm{m}},
\\
p_\mathrm{r} &= -\frac{2\, d_\mathrm{c} \, r_\mathrm{r}}{i_\mathrm{G}^2 \, J_\mathrm{c}  \, J_\mathrm{r}}.
\end{align*}
In transformed coordinates, the aerodynamic drag force $F_\mathrm{w}$ and the friction force $F_\mathrm{xr}$ can be written as 
\begin{gather*}
F_\mathrm{w} = 0.5 \, \rho_\mathrm{a} \, c_\mathrm{w} \, A_{\mathrm{st}} \, \eta_3 \, |\eta_3|, 
\\
F_{\mathrm{xr}} = \mu \, F_{\mathrm{zr}} \, \sin(C_\mathrm{r} \, \arctan(B_\mathrm{r} \, \lambda_{\mathrm{xr}}))
\shortintertext{with}
\nonumber\\[-6ex]
\lambda_{xr} = \frac{r_\mathrm{r}\, \eta_2 - \eta_3}{\mathrm{max}_\kappa \{
|r_\mathrm{r} \, \eta_2|_\kappa, |\eta_3|_\kappa 
\}}.
\end{gather*}  
Moreover, using the results of \cite{ReiOW2020}, it can be verified that the internal dynamics \eqref{eq:2WD_Byrnes_intern} are input-to-state stable with respect to the input $\xi$ for arbitrary vehicle parameters.
\subsection{Control Design}
\begin{figure*}[t]
\centering
\includegraphics[width=\columnwidth]{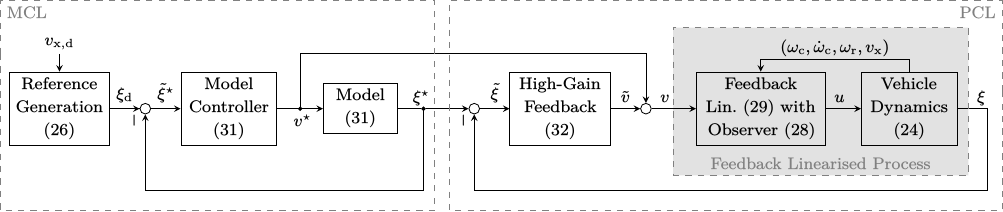}
\caption{Block diagram of the proposed high-gain model-following control design for advanced engine-based cruise control.}
\label{fig:example_mfc_blockdiagram}
\end{figure*}
The goal is to choose the desired motor torque $u = T_\mathrm{m,d}$ as the input for cruise control such that the vehicle velocity $v_\mathrm{x}$ tracks a reference $v_\mathrm{x,d}$.
The proposed control design is illustrated in Figure~\ref{fig:example_mfc_blockdiagram}.

Assuming that the reference considers gradual acceleration/deceleration with low load on the powertrain, we can indirectly enforce that $v_\mathrm{x}$ tracks $v_\mathrm{x,d}$ by reference tracking for the crankshaft angular velocity $\omega_\mathrm{c}$. 
In particular, due to the small twist angle $\phi_\mathrm{c}$, i.e. the assumption of a rigid drivetrain, and small slip $\lambda_\mathrm{xr}$, we have $	i_\mathrm{G}^{-1} \, \omega_\mathrm{c} \approx \omega_\mathrm{r}$ and $v_\mathrm{x} \approx r_\mathrm{r} \, \omega_\mathrm{r}$, i.e. the normalised crankshaft angular velocity matches the angular velocity of the wheels whose peripheral speed in turn matches the vehicle speed.
Therefore, we indirectly achieve tracking of the vehicle speed by designing the controller such that 
$y = i_\mathrm{G}^{-1} \, \omega_\mathrm{c}$ tracks
\begin{align}\label{eq:example_reference_xi}
y_\mathrm{d} =  \frac{v_\mathrm{x,d}}{r_\mathrm{r}}
\ \ \text{ with } \ \
\dot{y}_\mathrm{d} = \frac{\dot{v}_\mathrm{x,d}}{r_\mathrm{r}}
\ \ \text{ and } \ \
\ddot{y}_\mathrm{d} = \frac{\ddot{v}_\mathrm{x,d}}{r_\mathrm{r}}.
\end{align}
The measurements available for control are the wheel rotational velocity $\omega_\mathrm{r}$, the vehicle velocity~$v_\mathrm{x}$ as well as the normalised angular velocity $y$ and the normalised angular acceleration $\dot{y}$ of the crankshaft.
However, the twist angle $\eta_1 = \phi_\mathrm{c}$ is not available.
Moreover, the friction losses and the nonlinear tire model are uncertain.
Therefore, we cannot be apply feedback linearisation to render \eqref{eq:2WD_Byrnes_extern} an unperturbed integrator chain.
In particular, the uncertain terms $\tfrac{c_\mathrm{F}}{J_\mathrm{c}} (\tfrac{\xi_1}{\tau_\mathrm{m}} + \xi_2)$, $p_\mathrm{r}\, F_{\mathrm{xr}}$ and $p_\mathrm{c} \, \eta_1$ cannot be compensated.
In \cite{ReiOW2020a}, feedback linearisation is implemented with numerical derivatives to avoid the latter terms and a robust PID controller is used to cater for the friction losses. 
By adapting the implementation of \cite{ReiOW2020a}, however, we can formally guarantee stability in presence of the matched perturbations induced by the following.

\textbf{Friction:}
The coefficient $c_\mathrm{F}$ of friction is unknown. 
Therefore, we cannot compensate the term $ -\tfrac{c_\mathrm{F}}{J_\mathrm{c}} (\tfrac{\xi_1}{\tau_\mathrm{m}} + \xi_2)$.
In particular, we design the controller for the nominal case $c_\mathrm{F} = 0$ and consider $-\tfrac{c_\mathrm{F}}{J_\mathrm{c}} (\tfrac{\xi_1}{\tau_\mathrm{m}} + \xi_2)$ as perturbation.

\textbf{Tire model:}
Recall the friction term $F_{\mathrm{xr}}$ in \eqref{eq:Magic_Formular}.
The tire parameters $B_\mathrm{r}$, $C_\mathrm{r}$, and the friction coefficient $\mu$ of the road are uncertain.
In particular tire parameters $B_\mathrm{r}$ and $C_\mathrm{r}$ might depend on the vehicle configuration, mechanical wear, tire pressure, and environmental conditions, and $\mu$ is not measured and changes over time.
Thus, the term $p_\mathrm{r}\, F_{\mathrm{xr}}$ cannot be compensated.
In light of this, typically both the tire parameters and the friction coefficient are estimated. 
Given the estimates $\hat{B}_\mathrm{r}$, $\hat{C}_\mathrm{r}$ and $\hat{\mu}$ of $B_\mathrm{r},C_\mathrm{r}$ and $\mu$, an estimate of $F_{\mathrm{xr}}$ is
\begin{align}\label{eq:magic_formular_estimate}
\hat{F}_\mathrm{xr} = \hat{\mu} \, F_{\mathrm{zr}} \, \sin(\hat{C}_\mathrm{r} \, \arctan(\hat{B}_\mathrm{r} \, \lambda_{\mathrm{xr}})).
\end{align}
For convenience, we assume $\hat{B}_\mathrm{r}$, $\hat{C}_\mathrm{r}$ and $\hat{\mu}$ to be constant.
However, the estimates can also be dynamically obtained at run-time, \cite{IvaSS2015,AleV2017,RajPP2012,YiAH2002}.

\textbf{Twist angle:} 
The angle $\eta_1 = \phi_\mathrm{c}$ whose dynamics are given in \eqref{eq:example_model_twist_angle} is not measured.
Since both $y = i_\mathrm{G}^{-1} \, \omega_\mathrm{c}$ and $\omega_\mathrm{r}$ are measured, the trivial observer  
\begin{align}\label{eq:example_trivial_observer}
\dot{\hat{\eta}}_1 = i_\mathrm{G}^{-1} \, \omega_\mathrm{c} - \omega_\mathrm{r} = \xi_1 - \eta_2,
\end{align}
can be applied to estimate $\eta_1$.
That is, we integrate the difference of the angular velocities of the crankshaft and the wheels at run-time.
In doing so, the time derivative of the error $\eta_1 - \hat{\eta}_1$ is zero.
Thus, the error is given by 
$\eta_1(t) - \hat{\eta}_1(t) = \eta_1(0) - \hat{\eta}_1(0)$ for all $t \geq 0$.
In particular, for exact initialisation $\hat{\eta}_1(0) = \eta_1(0)$, we obtain an exact estimate $\hat{\eta}_1 \equiv \eta_1$.  

For the control design, we consider the desired external state $\xi_{\mathrm{d}} = \begin{bsmallmatrix}
y_\mathrm{d} \\ \dot{y}_\mathrm{d}
\end{bsmallmatrix}$.
Applying the feedback linearisation
\begin{gather}\label{eq:example_u_feedback_linearisation}
u = b^{-1}\big(
-a(\xi,\eta) +  \ddot{y}_\mathrm{d} + v
\big)
\\[-1ex]
\shortintertext{with}
\nonumber\\[-5ex]
a(\xi,\eta) = p_1  \, \xi_1 +  p_2  \, \xi_2  + p_3 \, \eta_2 + p_\mathrm{r} \, \hat{F}_\mathrm{xr} + p_\mathrm{c}  \, \hat{\eta}_1
\nonumber
\end{gather}
and $v(t) \in \mathbb{R}$, the dynamics of \eqref{eq:2WD_Byrnes}, \eqref{eq:example_u_feedback_linearisation} are given by
\begin{subequations}
\begin{align}
\dot{\xi} &= A \, \xi + B \big(
\ddot{y}_\mathrm{d} + v + \Delta(\xi,\eta)
\big),
\label{eq:example_closed_loop_extern}
\\
\dot{\eta} &= q(\xi,\eta)
\end{align}
\end{subequations}
with the perturbation
\begin{align*}
\Delta(\xi,\eta,t) \!=\!  
p_\mathrm{r} (F_\mathrm{xr} \!-\! \hat{F}_\mathrm{xr})
\! + \! p_\mathrm{c}  (\eta_1(0) \!-\! \hat{\eta}_1(0))
\!-\! \tfrac{c_\mathrm{F}}{J_\mathrm{c}} (\tfrac{\xi_1}{\tau_\mathrm{m}} \!+ \! \xi_2).
\end{align*}
Note that the dynamics are given as in \eqref{eq:system_feedback_linearisation}, and the internal dynamics are ISS.
Moreover, $F_\mathrm{xr}$ and $\hat{F}_\mathrm{xr}$ are bounded by construction in \eqref{eq:Magic_Formular} and \eqref{eq:magic_formular_estimate}.
Thus, $\Delta(\xi,\eta)$ satisfies \eqref{eq:perturbation_assumption} for some $\Delta,L_\Delta > 0$.
Consequently, we can apply the results of Section~\ref{sec:peaking}.
In particular, simulating the~model
\begin{gather}
\dot{\xi}^\star = A \, \xi^\star + B \big(
\ddot{y}_\mathrm{d} + v^\star
\big)
\ \ \text{with} \ \ 
\label{eq:example_model_mfc}
v^\star = -k^\star (\xi^\star - \xi_{\mathrm{d}})
\end{gather}
at run-time, we achieve arbitrarily good tracking precision with the control law \eqref{eq:example_u_feedback_linearisation} for $v = v^\star + \tilde{v}$ with
\begin{align}\label{eq:example_high_gain_feedback}
\tilde{v} = -\tilde{k}_\varepsilon^\top \, (\xi- \xi^\star)
\end{align}
and $\tilde{k}_\varepsilon$ from \eqref{eq:scaled_gain}.
Accordingly, the single-loop control~law 
\begin{align}\label{eq:example_u_sl}
u_\mathrm{sl} = b^{-1}\big(
-a(\xi,\eta)  +  \ddot{y}_\mathrm{d} - \tilde{k}_\varepsilon^\top (\xi- \xi_\mathrm{d})
\big),
\end{align}
which is obtained from~\eqref{eq:example_u_feedback_linearisation} for $v = -\tilde{k}_\varepsilon^\top \, (\xi- \xi_\mathrm{d})$, achieves the same precision.
In summary, we achieve arbitrarily good precision despite the uncertainties in both the friction losses and the tire model.

\begin{rem}
The uncertainty $\Delta$ depends on the state $(\xi,\eta)$ due to the friction losses and the uncertain tire model in $F_\mathrm{xr}$.
In particular, $\Delta(\xi,\eta)$ is, in general, not constant for nonconstant references $v_\mathrm{x,d}$. 
Thus, we cannot guarantee asymptotic tracking by augmenting the controller with an integral term.  
Furthermore, when considering the trivial estimate $\hat{F}_\mathrm{xr} = 0$ for the tire model, the entire term $F_\mathrm{xr}$ is considered as a perturbation, rather than the estimation error $F_\mathrm{xr} - \hat{F}_\mathrm{xr}$.   
\end{rem}

\begin{table}[htb]
	\centering
\begin{tabular}{clcc}
\textbf{Name} & \textbf{Description}  & \textbf{Value}  & \textbf{Unit}\\	
\hline 
\hline
$T_\mathrm{m}$ & \textnormal{Motor torque}   &  & \textnormal{N$\,$m} 
\\
$\omega_\mathrm{c},\xi_1$ &\textnormal{Crankshaft angular velocity} &  & \textnormal{$\frac{\text{rad}}{\text{s}}$}  
\\
$\dot{\omega}_\mathrm{c},\xi_2$ &\textnormal{Crankshaft angular acceleration} &  & \textnormal{$\frac{\text{rad}}{\text{s}^2}$}  
\\
$\phi_\mathrm{c},\eta_1$ & \textnormal{Twist angle torsional spring} &  & \textnormal{rad}  
\\
$\omega_\mathrm{r},\eta_2$ & \textnormal{Angular velocity of driveshafts} & & \textnormal{$\frac{\text{rad}}{\text{s}}$} 
\\
$v_\mathrm{x},\eta_3$ & \textnormal{Longitudinal vehicle velocity} & & \textnormal{$\frac{\text{m}}{\text{s}}$} 
\\
\hline
\hline
$\tau_\mathrm{m}$ & \textnormal{Motor time constant} & $0.02$ & \textnormal{s} 
\\
$i_\mathrm{G}$ & \textnormal{Total gear ratio} & $4.49$ & \textnormal{--} 
\\
$J_\mathrm{c}$ & \textnormal{Crankshaft moment of inertia} & $0.23$ &\textnormal{ kg$\,$m$^2$} 
\\
$c_\mathrm{F}$ & \textnormal{Coefficient of rational friction} & $0.003$ & $\frac{\textnormal{N} \, \textnormal{m}}{\textnormal{rad}/\textnormal{s}}$ 
\\
$k_\mathrm{c}$ & \textnormal{Stiffness torsional spring} & $5300$ &  $\frac{\textnormal{N} \, \textnormal{m}}{\textnormal{rad}}$ 
\\
$d_\mathrm{c}$ & \textnormal{Damping constant torsional spring} & $15$ & $\frac{\textnormal{N} \, \textnormal{m}\, \textnormal{s}}{\textnormal{rad}}$ \\
$J_\mathrm{r}$ &\textnormal{Wheel moment of inertia} & $3$ & \textnormal{kg$\,$m$^2$} 
\\
$C_\mathrm{r}$ & \textnormal{Pacejka shape factor} & $1.8$ & \textnormal{--} 
\\
$B_\mathrm{r}$ & \textnormal{Pacejka stiffness factor} & $10.3$ & \textnormal{--} 
\\
$m$ & \textnormal{Vehicle mass} & $1950$ & \textnormal{kg} 
\\
$g$ & \textnormal{Gravitational constant} & $9.81$ & \textnormal{$\frac{\text{m}}{\text{s}^2}$}\\
$l_\mathrm{f}$ &\textnormal{Distance front axle to COG} & $1.3$ & \textnormal{m} 
\\
$l_\mathrm{r}$ & \textnormal{Distance rear axle to COG} & $1.4$ & \textnormal{m} 
\\
$\rho_\mathrm{a}$ & \textnormal{Air density} & $1.1$ & \textnormal{$\frac{\text{k} \, \text{g}}{\text{m}^3}$} 
\\
$r_\mathrm{r}$ & \textnormal{Wheel radius} & $0.33$ & \textnormal{m}  
\\
$c_\mathrm{w}$ & \textnormal{Aerodynamic drag coefficient} & $0.3$ & \textnormal{--} 
\\
$A_\mathrm{st}$ & \textnormal{Vehicle front surface} & $2.37$ & $\phantom{^2}$\textnormal{m}$^2$
\\
$\mu$ & \textnormal{Tire-road friction coefficient} & $(0,1]$ & \textnormal{--} 
\\
\end{tabular}
\caption{States and parameters of the vehicle model, \cite{ReiOW2020a}.
}
\vspace{-2ex}
\label{table:vehicle_parameters}
\end{table}
\subsection{Simulation}
Let $\hat{\mu} = 0.9$, $\hat{B}_r = 11$ and $\hat{C}_r = 1.9$ be the estimate of the parameter $\mu = 1$, $B_\mathrm{r} = 10.3$ and $C_\mathrm{r} = 1.8$, respectively, and $\kappa = 10^{-6}$ in \eqref{eq:Slip_Coefficient_Rear}.
Choose the gain $k = \begin{bsmallmatrix}
1 \\ 2
\end{bsmallmatrix}$ and the scaling $\varepsilon = 0.15$.
We assume that there is small load on the powertrain at $t = 0$.
Moreover, we choose an exact initialisation for the model \eqref{eq:example_model_mfc} and the trivial observer~\eqref{eq:example_trivial_observer}, i.e.
\begin{align*}
\xi_0^\star = \xi_0 = \begin{bmatrix}
\tfrac{v_\mathrm{x}(0)}{r_\mathrm{r}} \ \, 0
\end{bmatrix}^\top \!\!\!,
\ 
\eta_0 = \begin{bmatrix}
0 \ \,	\tfrac{v_\mathrm{x}(0)}{r_\mathrm{r}} \ \, v_\mathrm{x}(0)
\end{bmatrix}^\top\!\!\!,
\
\hat{\eta}_1(0) = 0,
\end{align*}
with initial vehicle velocity $v_\mathrm{x}(0)$.
Inspired by \cite{ReiOW2020a}, we consider acceleration and deceleration with ramp-like references to benchmark the design.
Other scenarios of advanced cruise can be obtained by a combination of both.
In particular, we focus on the case of a reference $v_\mathrm{x,d}$ with inconsistent initial value, i.e. $v_\mathrm{x,d}(0) \neq v_\mathrm{x}(0)$, to demonstrate the peaking attenuation capability of the MFC in comparison the single-loop design.
Regarding the generation of $v_\mathrm{x,d}$ and $y_\mathrm{d}$, we consider the smooth~transition
\begin{align}\label{eq:example_smoothstep}
v_\mathrm{x,d}(t) = v_\mathrm{x,d}(0) + \big(v_\mathrm{x,d}(T) \!-\! v_\mathrm{x,d}(0)\big)  \big(3  \tfrac{t^2}{T^2}  - 2   \tfrac{t^3}{T^3}\big)
\end{align}
from $v_\mathrm{x,d}(0)$ to $v_\mathrm{x,d}(T)$ in time $T \geq 0$, and compute the corresponding $y_\mathrm{d}$ with \eqref{eq:example_reference_xi}. 

Regarding the implementation of the nonlinear MCL in Simulink, we note that the computational effort for simulating the efficient implementation \eqref{eq:example_model_mfc} is approximately 40\% lower than the effort for the classical implementation that considers the nonlinear model \eqref{eq:nonlinear_model}.

\begin{figure*}[ht]
\centering
\begin{subfigure}{0.49\linewidth}
\includegraphics[width=1\columnwidth]{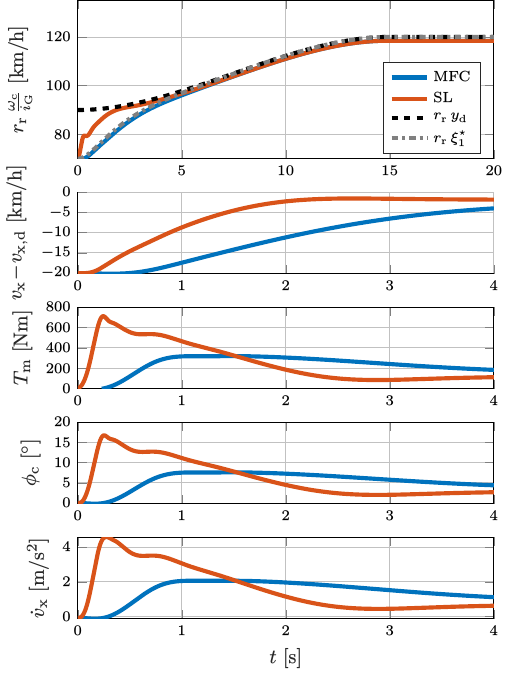}
\end{subfigure}
\begin{subfigure}{0.49\linewidth}
\includegraphics[width=1\columnwidth]{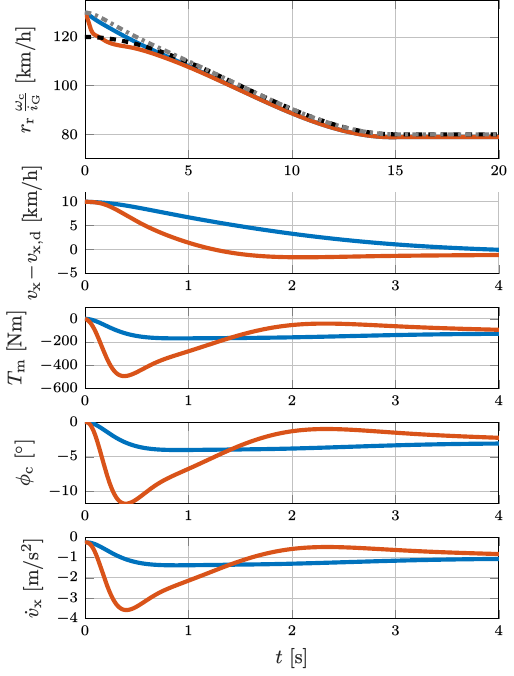}
\end{subfigure}
\caption{Model-following and single-loop tracking of a reference with inconsistent initial value: acceleration (left), deceleration~(right).}
\label{fig:example_benchmark}
\end{figure*}

\textbf{Acceleration benchmark:}
The goal is to accelerate from $70$ to $120$ km/h by tracking the (inconsistent) transition \eqref{eq:example_smoothstep} that interpolates between $90$ and $120$ km/h in $15$ seconds.
Afterwards, the velocity shall be constant. 
The plots in the left column of Figure~\ref{fig:example_benchmark} shows the acceleration whereas the right column shows the deceleration scenario.
The graphs show from top to bottom: the rotational velocity $\omega_\mathrm{c}$ of the crankshaft, the velocity deviation $v_\mathrm{x} - v_\mathrm{x,d}$, the motor torque $T_\mathrm{m}$, the twist angle $\phi_\mathrm{c}$, and the vehicle acceleration $\dot{v}_\mathrm{x}$ in closed-loop (MFC: blue line, single-loop: red line).
In the top plot, the desired crankshaft velocity $y_\mathrm{d}$ from \eqref{eq:example_reference_xi}, and the normalised solution $\xi_1^\star$ of the model are a dashed and a dash-dotted line, respectively.
For convenience, $\omega_\mathrm{c}$, $y_\mathrm{d}$ and $\xi^\star$ are normalised to the vehicle velocity which would be obtained for no slip and a rigid drivetrain, respectively.
Note that the range from $80$ to $110$ km/h corresponds to the rpm range from $3000$ to $4000$, and that the actual vehicle speed $v_\mathrm{x}$ is not equivalent to the normalised crankshaft velocity due to the tire slip and the flexible drivetrain.

The top plot shows that the normalised crankshaft velocity $\frac{\omega_\mathrm{c}}{i_\mathrm{G}}$ tracks the solution $\xi^\star_1$ of the model which in turn tracks the reference $y_\mathrm{d}$.
And with small slip, this yields a small ultimate bound for the velocity tracking error $v_\mathrm{x} - v_\mathrm{x,d}$, as shown in the second plot, i.e. the tracking requirement is satisfied.
Even though the MCL achieves exact tracking $\xi_1^\star \rightarrow y_\mathrm{d}$, the error between $\frac{\omega_\mathrm{c}}{i_\mathrm{G}}$ and $y_\mathrm{d}$ does not vanish in the PCL due to the uncertainty $\Delta$.
Note that, despite the inconsistency $y_\mathrm{d}(0) \neq \frac{\omega_\mathrm{c}(0)}{i_\mathrm{G}}$, i.e. $v_\mathrm{x,d}(0) \neq v_\mathrm{x}(0)$, the motor torque $T_\mathrm{m}$ does not peak since the MFC tracks the solution $\xi^\star$ of the model, which satisfies $\xi_0^\star = \xi_0$ by design.

\textbf{Deceleration benchmark:} The plots in the right column of Figure~\ref{fig:example_benchmark} show the deceleration from $130$ to $90$ km/h with the reference \eqref{eq:example_smoothstep} that interpolates between $120$ and $90$ km/h.
The results are similar to the acceleration case.
Note that the negative torque $T_\mathrm{m}$ of the hybrid/electric engine can be interpreted as regenerative breaking, \cite{LupDO2022}.

\begin{figure*}[ht]
\centering
\includegraphics[width=\columnwidth]{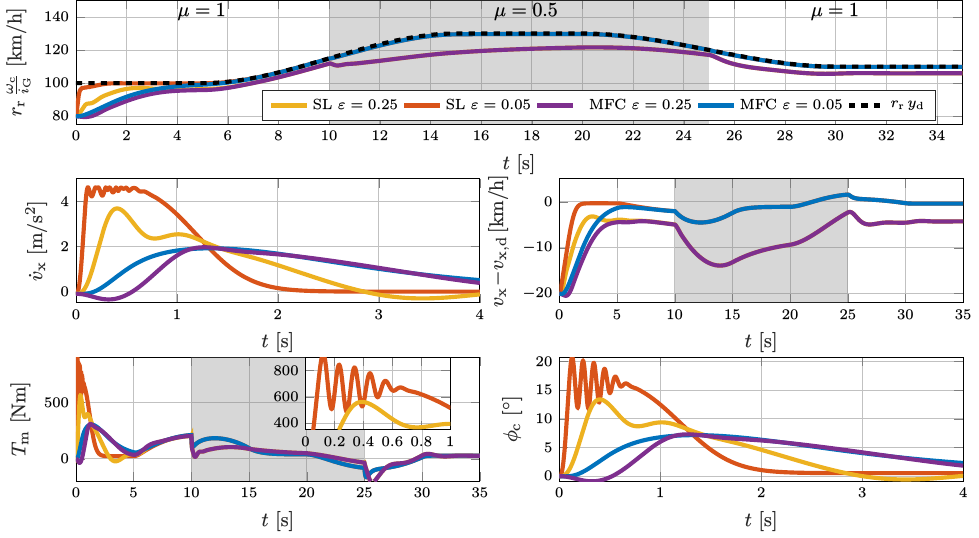}
\caption{Model-following and single-loop tracking of a reference with inconsistent initial value for changing road conditions.}
\label{fig:example_advanced_cruise}
\end{figure*}

For comparison, we consider the results (red lines) obtained for the single-loop control architecture consisting of the process \eqref{eq:2WD_Byrnes} and the controller \eqref{eq:example_u_sl}.
As expected, the design achieves the same tracking precision.
However, due to the inconsistent initial value of the reference, the high-gain feedback of the tracking error results in a large motor torque within the first second of the simulation.
This corresponds to undesirable behaviour, namely hard acceleration/breaking, and a lager (maximal) twist angle $\phi_\mathrm{c}$, resulting in higher stress and faster wearing of the power-train. 

Essentially, the advantage of the MFC is that we generate a reference $\xi^\star$, whose initial value is consistent to the initial state of the process by design.

\textbf{Advanced cruise:}
The acceleration and the deceleration references for the vehicle velocity can be combined for advanced manoeuvrers such as overtaking.
For overtaking, the vehicle velocity first increases, then remains constant, and finally decreases again. 
Figure~\ref{fig:example_advanced_cruise} shows such manoeuvre for a typical reference, whose initial value $v_\mathrm{x,d}(0)\neq v_\mathrm{x}(0)$ subject to changing road conditions.
In particular, the tire-road friction coefficient $\mu$ changes from $\mu = 1$ to $\mu = 0.5$ during the acceleration (at 10 sec.), and back to $\mu = 1$ during the deceleration (at 25 sec.).
Accordingly, the estimate $\hat{\mu}$ changes from $\hat{\mu} = 0.9$ to $\hat{\mu} = 0.4$ and back.
The time interval with $\mu = 0.5$ is highlighted by the grey shaded area.
To illustrate the influence of the scaling $\varepsilon$, the MFC (blue and purple lines) and the single-loop (red and yellow lines) design are shown for  $\varepsilon=0.25$ and $\varepsilon=0.05$, respectively.
After an initial transient of approximately 5 seconds the results of the single-loop and the MFC are very similar for the same scaling $\varepsilon$, i.e.
the red and the yellow lines align with the blue and the purple lines, respectively.

Two aspects are of particular interest, namely the influence of $\varepsilon$ and the peaking of the single-loop.
On the one hand, the ultimate bound for the tracking error $v_\mathrm{x} - v_\mathrm{x,d}$ can be rendered arbitrarily small by decreasing $\varepsilon$.
Note that the change of the road conditions at $10$ and $25$ seconds results in a large tracking error for the large scaling.
In comparison, the small scaling yields good tracking performance despite the change of $\mu$.
On the other hand, the single-loop requires larger motor torques than the MFC, i.e. $T_\mathrm{m}$ peaks for the single-loop, see bottom left plot.
Decreasing $\varepsilon$ to achieve good precision the peak of $T_\mathrm{m}$ becomes more pronounced for the single-loop.
The peaking results in an uncomfortable acceleration that exceeds the valid operating range of the model, and an undesirable oscillation in the power-train, in particular in $T_\mathrm{m}$ and $\phi_\mathrm{c}$.
The MFC design does not exhibit such peaking, in fact there is little change of $T_\mathrm{m}$ in the initial transient for different values of $\varepsilon$.
Moreover, the smaller scaling $\varepsilon$ (blue line) shows less aggressive control $T_\mathrm{m}$ for the changing road conditions while maintaining better tracking precision.


\subsection{Comparison to output-feedback with PI controller}
\begin{figure*}[th]
	\centering
	\includegraphics[width=\columnwidth]{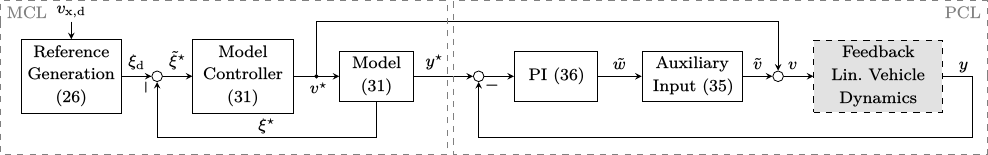}
	\caption{Block diagram of the proposed PI-based model-following control design for advanced engine-based cruise control.}
	\label{fig:example_mfc_blockdiagram_pid}
\end{figure*} 
In the following we investigate the potential of the MFC scheme using output-feedback with a PI-based design.
For the MFC we still apply the state feedback $v^\star$ within the MCL, but choose the auxiliary input $\tilde{v}$ as PI control rather than high-gain feedback in the PCL, see Figure \ref{fig:example_mfc_blockdiagram_pid}.

Given the feedback linearisation \eqref{eq:example_u_feedback_linearisation} with $v = v^\star + \tilde{v}$ and the model \eqref{eq:example_model_mfc}, the dynamics of the PCL are given by the perturbed integrator chain $\dot{\tilde{\xi}} = A \, \tilde{\xi} + B \, (\tilde{v} + \Delta(\xi,\eta))$ with the state $\tilde{\xi} = [\tilde{\xi}_1,\tilde{\xi}_2]^\top = \xi - \xi^\star$, whose first component $\tilde{\xi}_1= y - y^\star
= \xi_1 - \xi_1^\star$ is the error between the output of the process and the model.
Following \cite{ReiOW2020a}, we apply
\begin{align}\label{eq:example_reference_model}
	\tilde{v} = - a_0 \, \tilde{\xi}_1 - a_1 \, \tilde{\xi}_2 + b_0 \, \tilde{w}
\end{align}
with the auxiliary input $\tilde{w}(t) \in \mathbb{R}$. This yields
\begin{align*}
	\dot{\tilde{\xi}} = A \, \tilde{\xi} + B \big(
		- a_0 \, \tilde{\xi}_1 - a_1 \, \tilde{\xi}_2 + b_0 \, \tilde{w}(t) + \Delta(\xi,\eta)
	\big),
\end{align*}
where the choice $a_0 = 0$, $a_1 = \tau_\mathrm{m}^{-1}$, and $b_0 = i_\mathrm{G}^{-1} \, \tau_\mathrm{m} \, J_\mathrm{r}$ is motivated by the model of the drivetrain.
Choosing the gains $k_\mathrm{p} = 0.65$ and $k_\mathrm{i} = 0.16$ for the PI controller
\begin{align}\label{eq:example_pi_controller}
	\tilde{w}(t) = k_\mathrm{p}  (y^\star(t) - y(t)) + 	k_\mathrm{i} \textstyle \int_{0}^{t} (y^\star(\tau) - y(\tau)) \, \mathrm{d}\tau,
\end{align}
we obtain a phase margin of $75^\circ$ at the crossover frequency $\omega = 1$ rad/s for the open loop.

For comparison we consider a single-loop PI design with the auxiliary input 
$v = - a_0(\xi_1 \!-\! y_\mathrm{d}) - a_1(\xi_2 \!-\! \dot{y}_\mathrm{d})  + b_0 w$ 
and PI controller 
$w = k_\mathrm{p}(y_\mathrm{d} \!-\! y) + k_\mathrm{i}\int(y_\mathrm{y} \!-\! y) \, \mathrm{d}\tau$
for the feedback linearisation~\eqref{eq:example_u_sl}.

\begin{figure*}[th]
	\centering
	\includegraphics[width=\columnwidth]{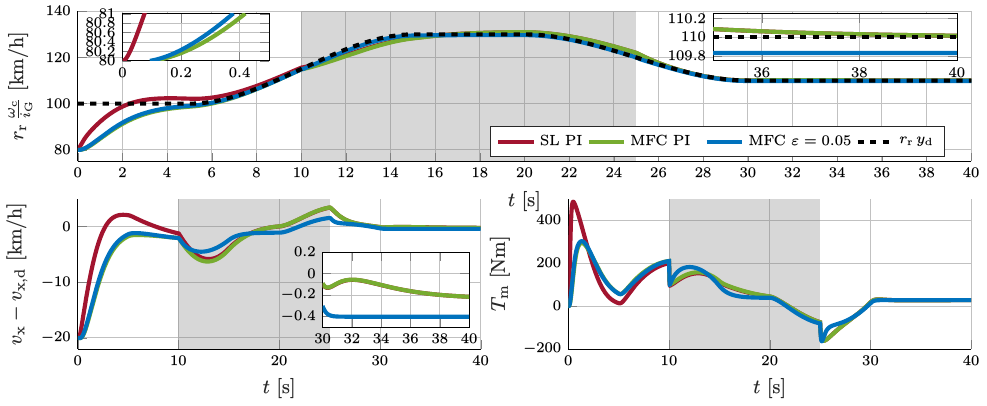}
	\caption{Model-following and single-loop tracking of a reference with inconsistent initial value with PI control design.}
	\label{fig:example_pid_comparison}
\end{figure*}
Figure~\ref{fig:example_pid_comparison} shows the results of the closed loop for the PI-based designs (dark red: single loop, green MFC PI) and the high-gain feedback MFC design with $\varepsilon=0.05$.
The tracking capabilities of the PI designs are comparable to that of the high-gain design. 
However, the PI based designs enforce a vanishing tracking error for the crankshaft velocity for constant speeds as shown in the top right zoom.
For constant references $y_\mathrm{d}$, the state $(\xi,\eta)$ converges to its stationary value.
Thus, the uncertainty $\Delta(\xi,\eta)$ converges to a constant value.
Consequently, the integral action, which cannot compensate $\Delta(\xi,\eta)$ for nonconstant references $\xi_{\mathrm{d}}$, enforces a vanishing tracking error during cruise.
Note, however, that the error $v_\mathrm{x} - v_\mathrm{x,d}$ still does not vanish completely due to the non-rigidity of the drivetrain. 
The single-loop PI design exhibits similar motor torque peaks as the single-loop high-gain design due to the inconsistent initial reference.

In summary we note that the peaking phenomenon is not exclusive to the high-gain state-feedback, but can also occur within classical output-feedback designs.
We design the engine-based cruise control using output-feedback for both, the single-loop and the MFC architecture.
It turns out that high gains are still required to achieve good tracking precision with the PI controller resulting in peaking of the single-loop control.
The comparable output-feedback using the MFC scheme provides the peaking attenuation advantages studied for the high-gain state-feedback in the previous sections.

\section{Conclusion}\label{sec:conclusion}
We study trajectory tracking using feedback linearisation and high-gain feedback within the model-following control architecture.
We present an efficient implementation of the controller, which reduces the computational effort at run-time, and we establish the capability of the MFC architecture to avoid peaking by leveraging the initial state of the model as an additional design parameter.
In particular, exact model initialisation decouples the transient response and the tracking precision of the high-gain design in the sense that the transient response is determined by the MCL, and the high-gain enforces the desired precision without dominating the transient.
The case study demonstrates the control design together with its the peaking attenuation capabilities and implementation efficiency.
For the studied engine-based cruise control, the peaking of the control signal results in undesirable acceleration profiles and fast wearing of the power-train. 
The case study demonstrates that both can be avoided by the proposed MFC scheme without sacrificing performance and robustness.

\vspace{2ex}
\setlength{\parindent}{0cm}

\textbf{\textsf{Research Funding:\hspace{0.5em}}}
This research is part of the project ``Model-following control for the trajectory-tracking control of nonlinear systems''
funded by the German Research Foundation (DFG), project number  508065537.

\section*{Appendix}
\textbf{Proof of Theorem~\ref{thm:tracking}:}
The dynamics of the closed-loop MFC system \eqref{eq:system}, \eqref{eq:u_feedback_linearisation}, \eqref{eq:linear_model} are given by
\begin{subequations}\label{eq:closed_loop}
\begin{align}
\dot{\tilde{\xi}}^\star &= (A - B \, k^\top) \, \tilde{\xi}^\star,
\label{eq:closed_loop_mcl}
\\
\tilde{\xi}\phantom{^\star} &= (A - B \, \tilde{k}^\top_\varepsilon) \, \tilde{\xi} + B \, \Delta(\xi,\eta,t), 
\label{eq:closed_loop_pcl_extern}
\\
\dot{\eta}\phantom{^\star} &= q(\xi,\eta).
\label{eq:closed_loop_pcl_intern}
\end{align}
\end{subequations}
Define the scaled state $\tilde{\zeta} \coloneqq D_\varepsilon^{-1} \tilde{\xi}$ with scaling~matrix
\begin{align*}
D_\varepsilon \coloneqq \mathrm{diag}\big(
\varepsilon^{n_\xi - 1},\varepsilon^{n_\xi - 2},...,\varepsilon,1
\big).
\end{align*}
The scaled state is bounded by
\begin{align}\label{eq:norm_scaled_state}
\lambda_{\mathrm{min}}(D^{-1}_\varepsilon) \, \Vert \tilde{\xi} \Vert_2 
\leq \Vert \tilde{\zeta} \Vert_2 
\leq \lambda_{\mathrm{max}}(D^{-1}_\varepsilon) \, \Vert \tilde{\xi} \Vert_2,
\end{align}
where $\lambda_{\mathrm{min}}(D_\varepsilon^{-1}) = 1$ and $\lambda_{\mathrm{max}}(D_\varepsilon^{-1}) = \varepsilon^{-(n_\xi-1)}$ denote the minimal and the maximal eigenvalue of $D_\varepsilon^{-1}$, respectively.
Moreover, noting that $D^{-1}_\varepsilon \, B = B$ and $\varepsilon \, D^{-1}_\varepsilon \, A \, D_\varepsilon = A$, 
it is readily verified that the external dynamics \eqref{eq:closed_loop_pcl_extern} of the PCL can be written as 
\begin{align}\label{eq:closed_loop_pcl_extern_scaled}
\varepsilon \, \dot{\tilde{\zeta}}\phantom{^\star} &= (A-B \, k^\top) \, \tilde{\zeta} + \varepsilon \, B \, \Delta(\xi,\eta,t). 
\end{align}
Since $\xi_{\mathrm{d}}$ is bounded by assumption and the origin of \eqref{eq:closed_loop_mcl} is asymptotically stable by design, $\xi^\star = \xi_{\mathrm{d}} + \tilde{\xi}^\star$ is bounded.
Moreover, the solution $\eta$ of the ISS internal dynamics is bounded whenever $\xi = \xi^\star + \tilde{\xi}$ is bounded.
Thus, to proof the theorem it remains to be shown that i) $\tilde{\xi}$ is bounded and ii) $\xi - \xi_{\mathrm{d}}$ is ultimately bounded with ultimate bound~$r_\infty$.

\noindent
i) Boundedness:
Let $P = P^\top > 0
$ be the solution of 
the Lyapunov equation $( A - B \, k^\top)^\top P + P \, (A - B\, k^\top) = -I$ with identity matrix $I$. Consider the Lyapunov~function%
\begin{align}\label{eq:lyapunov_function}
V_\varepsilon(\tilde{\xi}) = \tilde{\xi}^{\, \top}  D^{-1}_\varepsilon \, P \, D^{-1}_\varepsilon \, \tilde{\xi},
\end{align}
which depends on $\varepsilon$.
Note that $V_\varepsilon(\tilde{\xi}) = \tilde{\zeta}^\top \, P \, \tilde{\zeta}$.
Thus, 
\begin{align}\label{eq:norm_lyapunov_function}
\lambda_{\mathrm{min}}(P) \, \Vert \tilde{\zeta} \Vert_2^2
\leq V_\varepsilon(\tilde{\xi})
\leq \lambda_{\mathrm{max}}(P) \, \Vert \tilde{\zeta} \Vert_2^2.
\end{align}
Considering \eqref{eq:closed_loop_pcl_extern_scaled} and introducing $c = 2 \, \Vert P B \Vert_2$, it is readily verified that the time derivative of $V_\varepsilon(\tilde{\xi})$ satisfies 
\begin{align*}
\dot{V}_\varepsilon
&= -\varepsilon^{-1} \tilde{\zeta}^\top \, \tilde{\zeta} + 2 \, \tilde{\zeta}^\top P B \, \Delta(\xi,\eta,t)
\\
&\leq - \varepsilon^{-1}  \Vert \tilde{\zeta} \Vert_2^2 + c \, \Vert \tilde{\zeta} \Vert_2 \, |\Delta(\xi,\eta,t)|
\end{align*}
along the solution $\tilde{\xi}$ of \eqref{eq:closed_loop_pcl_extern} for $c = 2 \, \Vert P B \Vert_2$.
Moreover, inequality \eqref{eq:perturbation_assumption} is satisfied by assumption, and we obtain  
\begin{align*}
\Vert \xi \Vert_2 
= \Vert \xi^\star + \tilde{\xi} \Vert_2
\leq \Vert \xi^\star \Vert_2 + \Vert \tilde{\xi} \Vert_2
\leq \Vert \xi^\star \Vert_2 + \Vert \tilde{\zeta} \Vert_2
\end{align*}
using the first inequality of \eqref{eq:norm_scaled_state}.
Thus,
\begin{align}\label{eq:lyapunov_function_time_derivative_estimate}
\dot{V}_\varepsilon \leq -\Vert \tilde{\zeta} \Vert_2 \big(
(
\varepsilon^{-1} \! -\! c \, L_\Delta
) \, \Vert \tilde{\zeta} \Vert_2
- c \, (\delta + L_\Delta \Vert \xi^\star \Vert_2 )
\big).
\end{align}
Note that $	\varepsilon^{-1} - c \, L_\Delta > 0$ by assumption on $\varepsilon$, and $\xi^\star$ is bounded by design.
Therefore, the time derivative is negative whenever $\Vert \tilde{\zeta} \Vert_2$ is sufficiently large. 
Consequently, both $\tilde{\zeta}$ and $\tilde{\xi} = D_\varepsilon \, \tilde{\zeta}$ are bounded. 

\noindent
ii) Ultimate bound:
Since $\sup_{t \geq 0} \Vert \xi_{\mathrm{d}}(t) \Vert_2 < r_\mathrm{d}$ by assumption, we can choose $r_0  < r_\mathrm{d}$ such that $\sup_{t \geq 0} \Vert \xi_{\mathrm{d}}(t) \Vert_2 < r_0$.
Given some $x > 0$, define
\begin{gather}
c_\varepsilon(x) \coloneqq \frac{c \, (\delta + L_\Delta \, x)}{\varepsilon^{-1} - c \, L_\Delta},
\label{eq:bound_state_pcl}
\\
\tilde{\Omega}_\varepsilon(x) \coloneqq
\Big\{
\tilde{\xi} \in \mathbb{R}^{n_\xi}
\, \big| \, 
V_\varepsilon(\tilde{\xi}) \leq \lambda_{\mathrm{max}}(P) \, 	c_\varepsilon^2(x)
\Big\}.
\nonumber
\end{gather}
In \eqref{eq:lyapunov_function_time_derivative_estimate} the derivative of $V_\varepsilon(\tilde{\xi})$ is negative at time $t \geq 0$~if%
\begin{align*}
\Vert \tilde{\zeta}(t) \Vert_2  > c_\varepsilon( \Vert \xi^\star(t) \Vert_2),
\end{align*} 
which is for all elements outside the set $\tilde{\Omega}_\varepsilon(\Vert \xi^\star(t) \Vert_2)$ due to \eqref{eq:norm_lyapunov_function}.
Moreover, since $\sup_{t \geq 0} \Vert \xi_{\mathrm{d}}(t) \Vert_2 < r_0$ by assumption and $\tilde{\xi}^\star$ asymptotically converges to zero by design, there exists some  $T > 0$ (dependent on $r_0$) such~that
\begin{align*}
\Vert \xi^\star(t) \Vert_2 
\leq  \Vert \xi_{\mathrm{d}}(t) + \tilde{\xi}^\star(t) \Vert_2
\leq 
\Vert \xi_{\mathrm{d}}(t) \Vert_2 + \Vert \tilde{\xi}^\star(t) \Vert_2
< r_0
\end{align*}
for all $t \geq T$.
Thus, for $t \geq T$, the derivative of $\dot{V}_\varepsilon$ is negative outside the set $\tilde{\Omega}_\varepsilon(r_0)$.
Consequently, $\dot{V}_\varepsilon$ is negative on the boundary of $\tilde{\Omega}_\varepsilon(r_\mathrm{d}) \supset \tilde{\Omega}_\varepsilon(r_0)$, which shows that the solution $\tilde{\xi}$ enters the set $\tilde{\Omega}_\varepsilon(r_\mathrm{d})$ in finite time.
Moreover, due to the first inequalities of \eqref{eq:norm_scaled_state} and \eqref{eq:norm_lyapunov_function}, 
all elements of $\tilde{\Omega}_\varepsilon(r_\mathrm{d})$ satisfy $\Vert \tilde{\xi} \Vert_2 \leq \Vert \tilde{\zeta} \Vert_2 \leq p \, c_\varepsilon(r_\mathrm{d})$ with
\begin{align}\label{eq:eigenvalue_ratio}
p =\sqrt{\lambda_{\mathrm{min}}^{-1}(P) \, \lambda_{\mathrm{max}}(P)},
\end{align}
i.e. $\tilde{\Omega}_\varepsilon(r_\mathrm{d})$ is contained in the ball with radius $p \, c_\varepsilon(r_\mathrm{d})$, and it is readily verified that choosing $\varepsilon$ as in \eqref{eq:scaling_precision} enforces $p \, c_\varepsilon(r_\mathrm{d}) < r_\infty$.
Thus, the solution $\tilde{\xi}$ is ultimately bounded with ultimate bound $p \, c_\varepsilon(r_\mathrm{d}) < r_\infty$.
Moreover, the tracking error $\xi - \xi_{\mathrm{d}}$ can be written as in \eqref{eq:tracking_error}, where $\tilde{\xi}^\star$ asymptotically converges to zero by design.
Finally, $\xi - \xi_{\mathrm{d}}$ is ultimately bounded with the ultimate bound $r_\infty$, which concludes the proof. \\

\noindent
\textbf{Proof of Theorem~\ref{thm:u_comparison}:}
In order to compare the maximal control effort, first note that a lower bound of the supremum of $|u_\mathrm{sl}(t)|$ is given by the initial value $|u_\mathrm{sl}(0)|$, i.e. $\sup_{t \geq 0} |u_\mathrm{sl}(t)| \geq |u_\mathrm{sl}(0)|$.
Hence, we need to show that 
\begin{align}\label{eq:control_effort_comparison_proof}
\max_{(\xi_0,\eta_0) \in \mathcal{D} }\Big(
\sup_{t \geq 0} |u(t)|
\Big)
< 
\max_{(\xi_0,\eta_0) \in \mathcal{D} } |u_\mathrm{sl}(0)|.
\end{align}
We first i) compute a lower bound of the maximum of $|u_\mathrm{sl}(0)|$ for $(\xi_0,\eta_0) \in \mathcal{D}$, and then ii) overestimate the maximal value of $|u(t)|$ by considering the invariance properties of the MFC system.
Finally, comparing the bounds, we show that iii) there exists some $\bar{\varepsilon} $ such that \eqref{eq:control_effort_comparison_proof} is satisfied for all $\varepsilon < \bar{\varepsilon}$.
Denoting the ball of dimension $n$ with radius $r > 0$ centred at $x \in \mathbb{R}^n$ by $\mathcal{B}_r^{n}(x)$, both i) and ii) leverage that there exists $r,\, R > 0$ such that $\mathcal{B}_r^{n_\xi}(\xi_{\mathrm{d}}(0)) \subseteq \mathcal{D}_{\xi_0} \subseteq \mathcal{B}_R^{n_\xi}(\xi_{\mathrm{d}}(0))$ and $\mathcal{B}_r^{n_\eta}(0) \subseteq \mathcal{D}_{\eta_0} \subseteq \mathcal{B}_R^{n_\eta}(0)$ for the compact sets $\mathcal{D}_{\xi_0}$, $\mathcal{D}_{\eta_0}$.

\noindent
i) Lower bound of $|u_\mathrm{sl}(0)|$:
Applying the (reverse) triangle inequality to the initial value $u_\mathrm{sl}(0)$ from \eqref{eq:u_sl_initial_value}, we obtain
\begin{align*}
|u_\mathrm{sl}(0)| \geq \frac{|\tilde{k}_\varepsilon^\top  (\xi_0 - \xi_{\mathrm{d}}(0))| - |a(\xi_0,\eta_0)| - |y_\mathrm{d}^{(n_\xi)}(0) |}{|b(\xi_0,\eta_0)|},
\end{align*}
where $y_\mathrm{d}^{(n_\xi)}$ is bounded, and $a, \, b$ are continuous, and thus locally bounded.
Thus, there exist $\alpha \geq 0$, $\beta > 0$ such~that%
\begin{align*}
|u_\mathrm{sl}(0)| \geq \beta^{-1}\big(
|\tilde{k}_\varepsilon^\top \, (\xi_0 - \xi_{\mathrm{d}}(0))| - \alpha
\big)
\text{ for all }  (\xi_0,\eta_0) \in \mathcal{D}.
\end{align*}
Moreover, for each $\varepsilon > 0$ there exists some $\xi_0$ on the boundary of $\mathcal{B}_r^{n_\xi}(\xi_{\mathrm{d}}(0)) \subseteq \mathcal{D}_{\xi_0}$ such that $\xi_0 - \xi_{\mathrm{d}}(0)$ is parallel to $\tilde{k}_\varepsilon$.
By construction, this $\xi_0$ satisfies
\begin{gather}
|\tilde{k}_\varepsilon^\top \, (\xi_0 - \xi_{\mathrm{d}}(0))|
= \Vert \tilde{k}_\varepsilon \Vert_2 \, \Vert \xi_0 - \xi_{\mathrm{d}}(0) \Vert_2
= \Vert \tilde{k}_\varepsilon \Vert_2 \, r.
\nonumber
\shortintertext{Thus,}
|u_\mathrm{sl}(0)| \geq \beta^{-1}\big(
\Vert \tilde{k}_\varepsilon \Vert_2 \, r - \alpha
\big) 
\ \text{ for all } \ \eta_0 \in \mathcal{D}_{\eta_0}.
\label{eq:u_single_loop_peaking}
\end{gather}

\noindent
ii) Upper bound of $|u(t)|$: Applying the triangle inequality to \eqref{eq:u_feedback_linearisation}, \eqref{eq:linear_model}, and using $|b(\xi,\eta)| \geq b_\mathrm{m}$, we~obtain
\begin{align}
|u| \leq b_\mathrm{m}^{-1}\big(
|a(\xi,\eta)|+ |y_\mathrm{d}^{(n_\xi)}| + |k^\top \,  \tilde{\xi}^\star| + |\tilde{k}_\varepsilon^\top \, \tilde{\xi}|
\big),
\label{eq:u_mfc_inequality_u}
\shortintertext{where}
|k^\top \,  \tilde{\xi}^\star|
\leq \Vert k \Vert_2 \, \Vert \tilde{\xi}^\star \Vert_2
\quad \text{and} \quad
|\tilde{k}_\varepsilon^\top \, \tilde{\xi}| \leq \Vert \tilde{k}_\varepsilon \Vert_2 \, \Vert \tilde{\xi} \Vert_2.
\label{eq:u_mfc_inequality_u_linear_terms}
\end{align}
Next, we establish bounds of the solutions $\xi^\star$, $\xi$ and $\eta$.

a) Bound of $\xi^\star$:
The dynamics \eqref{eq:closed_loop_mcl} of $\tilde{\xi}^\star$ are asymptotically stable by design, and
\begin{align*}
\Vert \tilde{\xi}^\star_0 \Vert_2 
= \Vert \xi_0^\star - \xi_{\mathrm{d}}(0) \Vert_2
= \Vert \xi_0 - \xi_{\mathrm{d}}(0) \Vert_2
\leq R
\end{align*}
by assumption.
Thus, there exists $\tilde{r}^\star,r^\star > 0$ such that
\begin{align*}
\Vert \tilde{\xi}^\star(t) \Vert_2 \leq \tilde{r}^\star
\quad \text{and} \quad
\Vert \xi^\star(t) \Vert_2 
= \Vert \xi_{\mathrm{d}}(t) + \tilde{\xi}^\star(t) \Vert_2
\leq r^\star 
\end{align*}
for all $t \geq 0$.
In particular, using the Lyapunov function $V^\star(\tilde{\xi}^\star) = \tilde{\xi}^{\star \, \top} \, P \, \tilde{\xi}^\star$ 
it is readily verified that $\tilde{r}^\star$ can be chosen as $p \, R$ with $p$ from \eqref{eq:eigenvalue_ratio}.
Moreover, $\Vert \xi^\star \Vert_2  \leq \Vert \xi_{\mathrm{d}} \Vert_2 + \Vert \tilde{\xi}^\star \Vert_2$, and $\Vert \xi_{\mathrm{d}} \Vert_2$ is bounded by $r_\mathrm{d}$.
Thus, $r^\star = r_\mathrm{d} + p \, R$.

b) Bound of $\xi = \xi^\star + \tilde{\xi}$:
As shown in the proof of Theorem~\ref{thm:tracking}, the time derivative of $V_\varepsilon(\tilde{\xi})$ is negative at time $t \geq 0$ outside the set $\tilde{\Omega}_\varepsilon(\Vert \xi^\star(t) \Vert_2)$.
Since $\Vert \xi^\star(t) \Vert_2 \leq r^\star$ for all $t \geq 0$, we thus have that $\dot{V}_\varepsilon$ is negative outside the set $\tilde{\Omega}_\varepsilon( r^\star)$, and $\tilde{\xi}(0) = \xi_0 - \xi^\star_0 = 0  \in \tilde{\Omega}_\varepsilon(r^\star)$ for all $\varepsilon > 0$.
Thus, $\tilde{\xi}$ remains in the set $\tilde{\Omega}_\varepsilon( r^\star)$, which is contained in the ball with radius $p \, c_\varepsilon(r^\star)$.
Consequently, for all $t \geq 0$,
\begin{align*}
\Vert \xi(t) \Vert_2 
= \Vert \xi^\star(t) + \tilde{\xi}(t) \Vert_2 
\leq \Vert \xi^\star(t) \Vert_2 + \Vert \tilde{\xi}(t) \Vert_2
\leq r_{\varepsilon}
\end{align*}
where $r_\varepsilon = r^\star + p \, c_\varepsilon(r^\star)$ strictly decreases to $r^\star$ as $\varepsilon \rightarrow 0$. 

c) Bound for $\eta$:
The internal dynamics \eqref{eq:system_internal} of the process are ISS by assumption, and the external state $\xi$, which is considered as the input of \eqref{eq:system_internal}, satisfies $\Vert \xi(t) \Vert_2 \leq r_\varepsilon$ for all $t \geq 0$.
Moreover, $r_\varepsilon$ decreases as $\varepsilon$ decreases, and $\Vert \eta_0 \Vert_2 \leq R$ by assumption.
Thus, there exists $R_\varepsilon > 0$ (dependent on $\varepsilon$, $r$, and $R$), which does not increase as $\varepsilon $ decreases, such that $\Vert \eta(t) \Vert_2 \leq R_\varepsilon$ for all $t \geq 0$.
In particular, we obtain $R_\varepsilon$ as follows.	
As shown in \cite{SonW1995}, there exists a continuously differentiable, positive definite Lyapunov function $V_\eta$ such~that
\begin{subequations}
\begin{gather*}
\alpha_1(\Vert \eta \Vert_2) \leq V_\eta(\eta) \leq \alpha_2(\Vert \eta \Vert_2), \\[-0.5ex]
\pdv{V_\eta}{\eta} \, q(\xi,\eta) \leq - \alpha_3(\Vert \eta \Vert_2)
\ \text{ for all } \ \Vert \eta \Vert_2 \geq \gamma(\Vert \xi \Vert_2)
\end{gather*}
\end{subequations}
with class $\mathcal{K}_\infty$ functions $\alpha_1$ and $\alpha_2$, and class $\mathcal{K}$ functions $\alpha_3$ and $\gamma$.
Since $\Vert \xi \Vert$ is bounded by $r_\varepsilon$, it is readily verified that the time derivative $\dot{V}_\eta = \pdv{V_\eta}{\eta} \ q(\xi,\eta)$ is negative at time $t \geq 0$ whenever $V_\eta(\eta(t)) \geq \alpha_2(\gamma(r_\varepsilon))$.
Thus,	
\begin{align*}
V_\eta(\eta(t)) \leq \max\big\{
\alpha_2(\gamma(r_\varepsilon)),V_\eta(\eta_0)
\big\}
\ \text{ for all } \ t \geq 0.
\end{align*}
Moreover, $V_\eta(\eta_0) \leq \alpha_2(\Vert \eta_0 \Vert_2)$ by construction and $\Vert \eta_0 \Vert_2 \leq R$ by assumption.
Thus,
\begin{align*}
V_\eta(\eta(t))
\! \leq \! \max\{
\alpha_2(\gamma(r_\varepsilon)),\alpha_2(R)
\}
\! = \! \alpha_2(\max\{ \gamma(r_\varepsilon),R \})
\end{align*}
for all $t \geq 0$.
Consequently, $R_\varepsilon$ can be chosen as
\begin{align*}
R_\varepsilon = \alpha_1^{-1}( \alpha_2(\max\{\gamma(r_\varepsilon),R \})),
\end{align*}
which does not increase as $\varepsilon$ decreases.

Given the bounds $\tilde{r}^\star$, $p \, c_\varepsilon(r^\star)$, $r_\varepsilon$, and $R_\varepsilon$ of $\tilde{\xi}^\star$, $\tilde{\xi}$, $\xi$, and $\eta$, respectively, let $R_\mathrm{d}$ be an upper bound of the bounded time derivative $|y_\mathrm{d}^{(n_\xi)}|$, consider the balls $\mathcal{B}_{r_\varepsilon}^{n_\xi}(0)$ and $\mathcal{B}_{R_\varepsilon}^{n_\eta}(0)$.
Considering \eqref{eq:u_mfc_inequality_u} and \eqref{eq:u_mfc_inequality_u_linear_terms}, we readily obtain 
\begin{gather}
|u| \leq b_\mathrm{m}^{-1}\big(
\rho(\varepsilon) + p \, c_\varepsilon(r^\star)\,  \Vert \tilde{k}_\varepsilon \Vert_2
\big)
\label{eq:u_mfc_upper_bound}
\\[-1.25ex]
\shortintertext{where}
\rho(\varepsilon) = R_\mathrm{d} + \Vert k \Vert_2 \, \tilde{r}^\star + \displaystyle \mathrm{max}_{
(\xi,\eta) \in \mathcal{B}_{r_\varepsilon}^{n_\xi}(0) \times \mathcal{B}_{R_\varepsilon}^{n_\eta}(0)
}|a(\xi,\eta)|
\nonumber
\end{gather}
cannot increase as $\varepsilon$ decreases since $r_\varepsilon$ and $R_\varepsilon$ cannot increase and $\tilde{r}^\star$ is independent of $\varepsilon$.

\noindent
iii) \eqref{eq:control_effort_comparison_proof} for $\varepsilon < \bar{\varepsilon}$:
Note that \eqref{eq:control_effort_comparison_proof} is satisfied if the right-hand side of the lower bound \eqref{eq:u_single_loop_peaking} of $|u_\mathrm{sl}(0)|$ is larger than the right-hand side of the upper bound \eqref{eq:u_mfc_upper_bound} of $|u|$, i.e.
\begin{align}\label{eq:comparison_u_inequality_max}
b_\mathrm{m}^{-1}\big(
\rho(\varepsilon) + p \, c_\varepsilon(r^\star)\,  \Vert 	\tilde{k}_\varepsilon \Vert_2
\big)
<
\beta^{-1}\big(
\Vert \tilde{k}_\varepsilon \Vert_2 \, r - \alpha
\big),
\end{align}
where $b_\mathrm{m}, \, \beta, \, p \, c_\varepsilon(r^\star) > 0$.
The inequality holds~whenever
\begin{align}\label{eq:comparison_u_inequality}
\Vert \tilde{k}_\varepsilon \Vert_2 \, \big(
b_\mathrm{m} \, r - \beta \, p \, c_\varepsilon(r^\star)
\big) - b_\mathrm{m} \, \alpha - \beta \, \rho(\varepsilon) > 0.
\end{align}
Since $c_\varepsilon(r^\star)$ continuously decreases to zero as $\varepsilon \rightarrow 0$, for each $\Theta \in (0,1)$, there exists $\varepsilon_r > 0$ such~that
\begin{align*}
b_\mathrm{m} \, r - \beta \, p \, c_\varepsilon(r^\star) > \Theta \, b_\mathrm{m} \, r
\quad \text{ for all } \quad \varepsilon < \varepsilon_r.
\end{align*}
Consequently, the left-hand side of \eqref{eq:comparison_u_inequality} is larger than
\begin{align}\label{eq:comparison_u_inequality_estiamte}
\Theta \, b_\mathrm{m} \, r \, \Vert \tilde{k}_\varepsilon \Vert_2 - b_\mathrm{m} \, \alpha - \beta \, \rho(\varepsilon)
\end{align}
for all $\varepsilon < \varepsilon_r$, where the norm $\Vert \tilde{k}_\varepsilon \Vert_2$ of the gain from \eqref{eq:scaled_gain} becomes arbitrarily large as $\varepsilon \rightarrow 0$, and $\rho(\varepsilon)$ cannot increase for decreasing $\varepsilon$.
Consequently, there exists some $\bar{\varepsilon} < \varepsilon_r$ such that \eqref{eq:comparison_u_inequality_estiamte} is positive for all $\varepsilon 
< \bar{\varepsilon}$.
Then, \eqref{eq:comparison_u_inequality} is satisfied.
Finally, inequality \eqref{eq:comparison_u_inequality_max} holds.

\input{AT_MFC_Cruise_Control_publisher.bbl}

\end{document}